\documentclass[%
reprint,
superscriptaddress,
nofootinbib,
twocolumn,
notitlepage,
amsmath,amssymb,
aps,
pra
]{revtex4-2}

\usepackage{graphicx}
\usepackage{dcolumn}
\usepackage{bm}
\usepackage{tablefootnote}

\usepackage{amsthm}
\usepackage{mathtools}
\usepackage{amsmath}
\usepackage{amssymb}
\usepackage{graphicx}
\usepackage{url}
\usepackage{float}
\usepackage{bm}
\usepackage{ifthen}
\usepackage[usenames,dvipsnames]{color}
\usepackage{mathrsfs}
\usepackage[colorlinks=true,citecolor=blue,urlcolor=black]{hyperref}
\usepackage{float}
\usepackage{subfigure}
\usepackage{enumitem}
\usepackage{color}
\usepackage{setspace}
\usepackage{braket}
\usepackage{comment}
\usepackage{lipsum}

\newcommand{\be}{\begin{equation}}
\newcommand{\ee}{\end{equation}}

\begin{document}

\title{Fully-Passive Quantum Key Distribution}



\author{Wenyuan Wang}
\thanks{wenyuanw@hku.hk}
\affiliation{Department of Physics, University of Hong Kong, Pokfulam Road, Hong Kong}

\author{Rong Wang}
\affiliation{Department of Physics, University of Hong Kong, Pokfulam Road, Hong Kong}

\author{Victor Zapatero}
\affiliation{Escuela de Ingeniería de Telecomunicación, Department of Signal Theory and Communications, University of Vigo, Vigo, Spain}
\affiliation{atlanTTic Research Center, University of Vigo, E-36310 Vigo, Spain}

\author{Li Qian}
\affiliation{Centre for Quantum Information and Quantum Control (CQIQC), Dept. of Electrical \& Computer Engineering, University of Toronto, Toronto,  Ontario, M5S 3G4, Canada}

\author{Bing Qi}
\affiliation{Cisco Systems, San Jose, CA 95134, USA}

\author{Marcos Curty}
\affiliation{Escuela de Ingeniería de Telecomunicación, Department of Signal Theory and Communications, University of Vigo, Vigo, Spain}
\affiliation{atlanTTic Research Center, University of Vigo, E-36310 Vigo, Spain}

\author{Hoi-Kwong Lo}
\thanks{hoikwong@hku.hk}
\affiliation{Department of Physics, University of Hong Kong, Pokfulam Road, Hong Kong}
\affiliation{Centre for Quantum Information and Quantum Control (CQIQC), Dept. of Electrical \& Computer Engineering, University of Toronto, Toronto,  Ontario, M5S 3G4, Canada}
\affiliation{Centre for Quantum Information and Quantum Control (CQIQC), Dept. of Physics, University of Toronto, Toronto,  Ontario, M5S 3G4, Canada}


\begin{abstract}
	
	Passive implementations of quantum key distribution (QKD) sources are highly desirable as they eliminate side-channels that active modulators might introduce. Up till now, passive decoy-state and passive encoding BB84 schemes have both been proposed. Nonetheless, passive decoy-state generation and passive encoding have never been simultaneously implemented with linear optical elements before, which greatly limits the practicality of such passive QKD schemes. In this work, we overcome this limitation and propose a fully-passive QKD source with linear optics that eliminates active modulators for both decoy-state choice and encoding. This allows for highly practical QKD systems that avoid side-channels from the source modulators. The passive source we propose (combined with the decoy-state analysis) can create any arbitrary state on a qubit system and is protocol-independent. That is, it can be used for various protocols such as BB84, reference-frame-independent QKD, or the six-state protocol. It can also in principle be combined with e.g. measurement-device-independent QKD, to build a system without side-channels in either detectors or modulators. 
	
\end{abstract}

\date{\today}
\maketitle

\section{Background}

Quantum key distribution (QKD) \cite{bb84} holds the promise of information-theoretically secure communications between two parties, Alice and Bob. However, while QKD is theoretically secure, practical components in a QKD system might contain side-channels, which might bring loopholes to the implementation security of such a QKD system \cite{blinding,timeshift,IMleak1,IMleak2,trojan,trojan2}.

Measurement-device-independent (MDI) QKD \cite{mdiqkd} has been proposed to eliminate all side-channels in \textit{the measurement unit}. The more recently proposed Twin-Field (TF) QKD \cite{TFQKD} protocol provides similar MDI property, while also offering better scaling of the key rate with respect to channel loss.

However, current QKD implementations provide relatively less protection against side-channels in the \textit{sources}, which are usually assumed to be trusted. In practice, modulators could inadvertently leak information during the encoding process. For instance, polarization modulators \cite{IMleak1} and intensity modulators \cite{IMleak2} can have side-channels that either directly leak information to Eve, or undermine the security of the decoy-state analysis. Also, Eve could even potentially send a ``Trojan Horse" \cite{trojan,trojan2} into the source that might leak information to Eve along with the actual signal. Therefore, a passive QKD setup, where there are no active modulators and the encoding is entirely performed via post-selection, would be highly desirable because it eliminates the side-channels from the source modulators.

To address this problem, passive decoys \cite{passivedecoy1,passivedecoy2} have been proposed to avoid side-channels from the intensity modulators encoding the decoy settings used in decoy-state analysis \cite{decoystate_Hwang,decoystate_LMC,decoystate_Wang} for weak coherent pulse (WCP) sources. The passive decoy scheme makes use of the idea that when two phase-randomized coherent pulses interfere at a beam-splitter, the state of the output at one port is correlated to that of the other port. This means that we can detect the signal at one output port, to conditionally determine intensity (or the photon number distribution) at the other port, hence categorizing signals into different ``decoy regions" during post-selection. There have also been multiple reports of experimental implementations of passive decoy states \cite{passive_experiment1,passive_experiment2,passive_experiment3}. The idea of using random phase fluctuations in the sources and interfering two pulses has also been applied successfully to quantum random number generators (QRNGs) \cite{QRNG1,QRNG2,QRNG3,phaserandom2}. 

Additionally, there have been proposals and also experimental demonstrations for implementing passive decoys with entanglement sources and local detectors (i.e. heralded single-photon source) \cite{passiveHSPS,passiveHSPSexperiment}, but these sources suffer from lower repetition rate compared to WCP sources. In this work we will focus on implementations with WCP sources only.

On the other hand, a passive encoding scheme \cite{passiveEncoding} for BB84 states has also been proposed. It makes use of post-selection, but rather than interfering two identical signals and post-selecting the output (where the phase difference between the two incoming pulses determine the output intensity at one port), here two incoming signals of different polarization (say H and V) superimpose at a polarizing beam-splitter. The output polarization is determined by the phase difference between incoming pulses, meaning that post-selection can be used to obtain the four encoding states for BB84. 

Passive-encoding has also been proposed for continuous-variable QKD protocols, such as in Ref. \cite{passiveCV}, while in this work we mainly focus on discrete-variable protocols (polarization or time-bin phase encoding) based on WCP sources. Also, another work worth noting is Ref. \cite{fourlaser} that makes use of four lasers driven by electrically-modulated signals to create different decoy settings, and then uses polarizers to generate BB84 signals, so no optical modulator is employed. However, the electrical signals still carry the basis bit and decoy setting information (and thus side-channels could still exist in the modulation). In this work when referring to ``passive" encoding and decoy generation, we mean that Alice decides her encoding and decoy setting via post-selection only. The main advantage of such a scheme is that it avoids the modulation-related side-channels, optical or electrical.\\

Up till now, though, for more than 10 years after its proposal, passive encoding has never been successfully combined with passive decoy generation in a simple setup with linear components and WCP sources. The main challenge is that: \textit{the intensity and polarization of the prepared states are correlated in a passive QKD setup}. This seems to make passive-encoding QKD incompatible with passive decoy state generation, thus severely limiting its practical use. 

It is worth noting that one previous work \cite{passiveBB84decoy} has included a scheme to implement simultaneously passive encoding and passive decoy preparation, but it makes use of sum-frequency generation, a nonlinear effect, which increases the difficulty of implementing it in a practical QKD system. In this work, we focus on addressing the challenge of implementing a passive QKD setup with linear components and WCP sources only, which, if successful, will tremendously improve the practicality of passive QKD setups.

We observe that the aforementioned challenge, i.e. the fact that the intensity and polarization are correlated, comes from two factors: (1) The biggest problem is that, in the passive encoding and passive decoy protocols so far, only 2 degrees-of-freedom (DOFs) are used, which come from the random phases of two laser sources. Since global randomization requires one DOF, this only leaves one DOF to simultaneously determine the intensity and polarization, which makes them inevitably correlated. (2) Secondly, even if a source can emit states of arbitrary combinations of intensity and polarization, the occurrences of such states would follow a classical distribution, and the probability distributions of intensity/polarization might be correlated. 

Since the post-selection regions in a passive scheme have a finite size (i.e. Alice cannot choose an exact state but can only post-select a range of states), this means that Alice needs to perform the decoy-state analysis on coherent states with \textit{mixed and correlated} intensities and polarizations, which could result in security loopholes, e.g. the average yield or QBER associated with an n-photon state may depend on the decoy-setting.

In this work, we make two main contributions: (1) We propose a passive source capable of creating a general-purpose source state of arbitrary polarization \textit{and} arbitrary intensity independently. This is achieved by using four laser sources at once, thus ensuring that enough DOFs are available in the source for the intensity, the polarization, and the global phase randomization. (2) We propose a set of carefully designed post-selection strategies to decouple the intensity and polarization distributions, thus enabling the use of the standard decoy-state analysis even when the post-selection regions are finite. This work enables \textit{fully-passive} QKD with both the encoding and the decoy state generation implemented passively via local detection and post-selection only. 

Importantly, the passive source we propose is general and protocol-independent since it can create any arbitrary polarization state on the Bloch sphere and arbitrary intensity. In practice, by choosing an appropriate post-selection strategy, it can for instance be used for BB84 \cite{bb84}, reference-frame-independent (RFI) QKD \cite{RFIQKD}, or the six-state protocol \cite{sixstate}. Moreover, the same source can in principle be combined with e.g. MDI-QKD \cite{mdiqkd} to create a QKD system without side-channels in either detectors or modulators. With our proposal, we are finally able to introduce a practical QKD scheme without the susceptibilities of source modulator side-channels, yet still maintaining acceptable key rates.

This paper is divided into two main topics: the source and the protocol. Firstly, in Section II we briefly review the passive decoy and passive encoding setups, and describe our new fully passive source that is capable of creating states with arbitrary polarization on the Bloch sphere and arbitrary intensity. Secondly, in Section III we describe a set of post-selection strategies to use our new passive source in various QKD protocols with two or three pairs of bases in $\{H,V,+,-,L,R\}$ polarizations. (Note that our source is general, and this is only one approach among many.) More specifically, we define post-selection areas in Sec. III.A and a new type of post-selection on the intensity distribution in Sec. III.B. We briefly discuss the security of using single-photons with mixed polarizations as signal states in Sec. III.C, and address the main challenge, i.e. the use of coherent states with \textit{mixed and correlated} polarizations and intensities to perform the decoy-state analysis, in Sec. III.D. We also discuss the practical challenges in implementing our scheme experimentally in Sec. III.E. We simulate the performance of two example protocols, BB84 and RFI-QKD, and also show the practicality of our scheme in the finite-size regime in Sec. IV. Lastly, we include some discussions and remarks in Sec. V.

Overall, we propose both a general-purpose fully passive source as well as a set of theoretical and numerical tools to use it effectively and securely in a QKD protocol. This enables a setup without side-channels from the source modulators, and can also potentially be used in a MDI protocol to enable even higher implementation security.

\section{Fully Passive Source}

\subsection{Passive Decoy and Passive Encoding}

\begin{figure}[h]
	\includegraphics[scale=0.225]{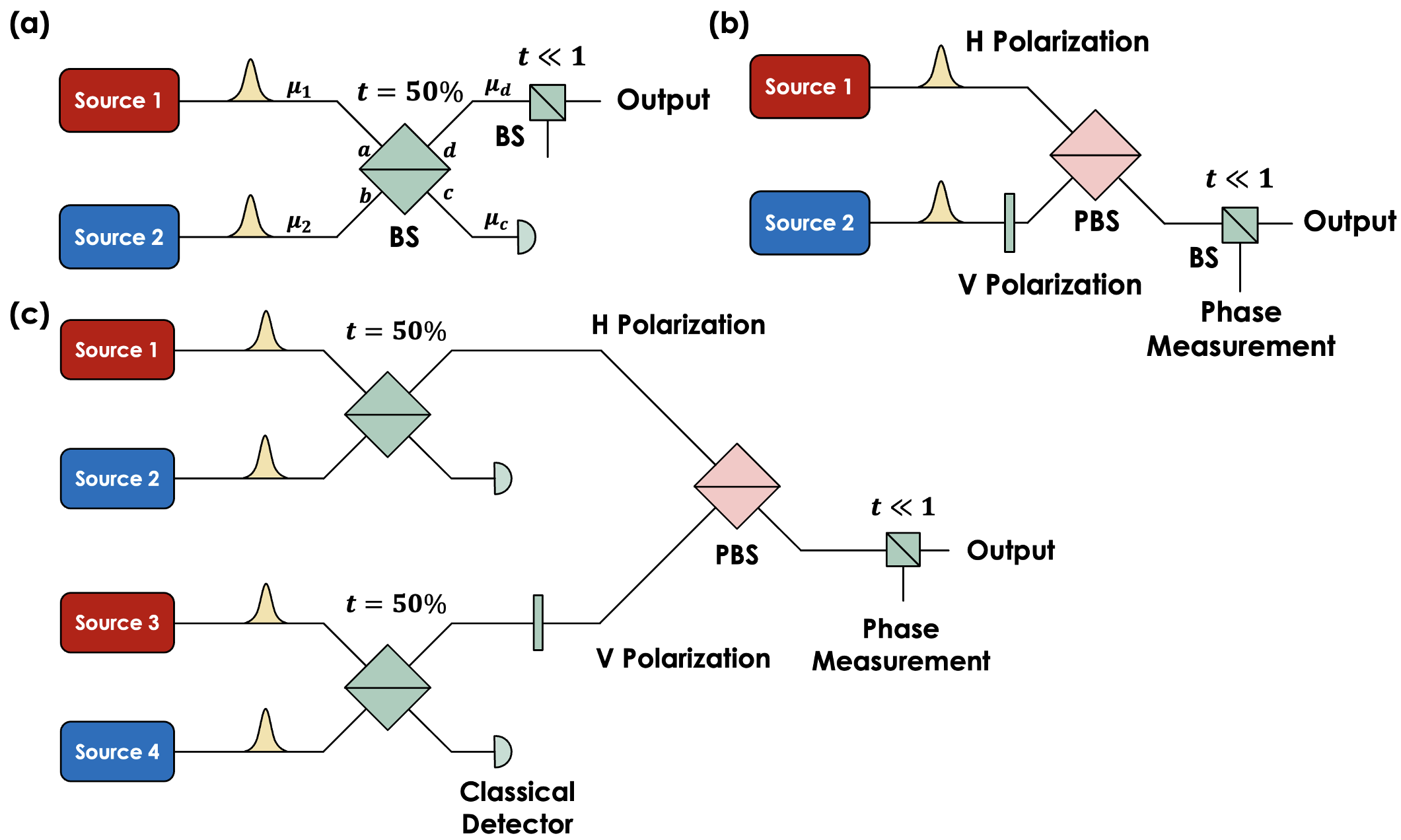}
	\caption{(a) Passive decoy-state setup. The phase difference between two sources determine the output intensity, while the global phase is randomized. (b) Passive BB84 encoding setup. The phase difference between two sources determine the output polarization, while the global phase is randomized. (c) Fully passive QKD source. There are a total of 4 degrees of freedom coming from the random phases of the four sources. These variables determine the intensities coming from the upper and the lower arms, and the phase difference and intensity mismatch between upper/lower arms determine the output polarization state, which can be any state on the Bloch sphere. Lastly, the output state has a random global phase. Alice can perform post-selection based on her two classical detectors (which can be photodiodes measuring intensities) and her measurement on the relative phase between the two polarization modes.}
	\label{fig:1}
\end{figure} 

Before we introduce our setup, let us first briefly review how passive decoy-state and passive BB84 encoding work.

Passive decoy-state was first proposed in Ref. \cite{passivedecoy1}, where it was observed that two phase-randomized coherent light beams interfering at a beam splitter yield intensity correlations at the two output ports. For any given input phases, the output for two interfering coherent light pulses (supposing a 50:50 beam splitter) is:

\begin{equation}
	\begin{aligned}
		&\ket{\sqrt{\mu_1} e^{i\phi_1}}_a \ket{\sqrt{\mu_2} e^{i\phi_2}}_b \rightarrow \\
		&\ket{\sqrt{\mu_1/2} e^{i\phi_1}+i\sqrt{\mu_2/2} e^{i\phi_2}}_c
		\ket{i\sqrt{\mu_1/2} e^{i\phi_1}+\sqrt{\mu_2/2} e^{i\phi_2}}_d
	\end{aligned}
\end{equation}

\noindent which is simply the product of two coherent output states. However, if the phase difference is randomized (i.e. if we integrate the above state over $\phi = \phi_1 - \phi_2$ in the range of $[0,2\pi)$), the photon numbers on the two output ports would be correlated, while still leaving the global phase randomized (and unknown to Eve, hence allowing the use of decoy states).

To make use of the correlation and post-select the output states, one option, as described in Ref. \cite{passivedecoy1}, is to use WCP sources in combination with a single-photon detector (SPD) to observe one output port, and post-select events into ``click" and ``no-click" bins, hence conditionally determining the state from the other port. 

The other perhaps conceptually even simpler option, as later described in Ref. \cite{passivedecoy2}, is to prepare $\mu_1$ and $\mu_2$ as strong lights, use a classical detector at one port for post-selection, and then use a beam splitter with small transmittance to reduce the intensity of the output signal to the single-photon level for use in QKD. In this case, one can deterministically determine the intensity $\mu_d$ at one output port of the beam splitter, which means the other output port has intensity $\mu_c = \mu_1 + \mu_2 - \mu_d$. One can simply use the intensity to divide signals into bins (e.g. with a threshold value $\mu_T$ and categorize signals into ``above threshold" versus ``below threshold" events). The signals in these bins will on average have a different photon number distribution. An example setup is illustrated in Fig. \ref{fig:1} (a).\\

Passive BB84 encoding was proposed in Ref. \cite{passiveEncoding}, and uses a very similar setup as passive decoys, except that the two input states have different polarizations H and V, and are combined at a polarizing beam splitter. The output state is still a coherent state:

\begin{equation}
	\begin{aligned}
		\ket{\psi} = \sum_{n=0}^{\infty} \sqrt{P_n} \ket{n}_{\phi_{HV}},
	\end{aligned}
\end{equation}

\noindent where $P_n=e^{-\mu}\mu^n/n!$ is the Poissonian distribution, the total output intensity is $\mu$, and the photon number state is

\begin{equation}
	\begin{aligned}
		\ket{n}_{\phi_{HV}} &= {1\over{\sqrt{n!}}}{a^{\dagger}}_{\phi_{HV}}^n \ket{vac},\\
		{a^{\dagger}}_{\phi_{HV}} &= (a_H^\dagger + e^{i\phi_{HV}} a_V^\dagger)/\sqrt{2},\\
	\end{aligned}
\end{equation}

\noindent which is a state that lies on the equator of the Bloch sphere. Here the polarization is entirely determined by the relative phase $\phi_{HV}$ between two polarization modes. By splitting off part of the strong classical signal and measuring its polarization, one can post-select those polarizations that are close to the four BB84 polarizations $+,-,L,R$. An example setup is illustrated in Fig. \ref{fig:1} (b). 

However, the main drawback is that up till now, passive encoding and passive decoy generation have not been successfully combined using only linear optical components. The reason is that, based on the above schemes, the same phase difference information is either used to determine the decoy setting (i.e. output intensity) or the polarization encoding. The two sources only contain two degrees-of-freedom each, which are respectively used for post-selection and for global phase randomization. Therefore, one can only implement one type of post-selection at a time, for either encoding or decoy-state.\\

\subsection{Scheme for Fully Passive Source}

Here we propose a simple yet innovative idea to set up a source that allows Alice to perform both passive encoding and decoy state preparation. The source itself is general and protocol-independent and can output a phase-randomized coherent state with arbitrary polarization and intensity. When combined with the decoy-state analysis, we can in principle prepare single photons in any qubit state. In practice, Alice can specify her post-selection strategy to create different desired states, such as the four BB84 states $\{H,V,+,-\}$, or six states $\{H,V,+,-,L,R\}$ for the reference-frame-independent or the six-state protocols.

In our setup, Alice uses four independent sources, which allows her to have four (rather than two) degrees-of-freedom. An illustration of our setup can be found in Fig. \ref{fig:1} (c). (1) Two pairs of sources first interfere at two beam splitters respectively, where a first step of post-selection is implemented by observing one output port of each beam splitter, resulting in two output pulses of arbitrary intensities depending on the post-selection results. (2) Then, these pulses are rotated into H and V polarizations respectively and are combined at a polarizing beam splitter (PBS). Depending on the intensities of the two incoming signals, the output state of the PBS can be of any intensity and any polarization on the Bloch sphere. (3) The output signal (at this point still being strong light) is split off to perform a polarization measurement, which constitutes a second step of post-selection, and meanwhile the output signal is attenuated to the single-photon level as the final prepared state.

The output state entirely depends on the four degrees-of-freedom in the system, namely the random phases $\{\phi_1,\phi_2,\phi_3,\phi_4\}$ of the four independent sources. Depending on the phase difference between the upper pair and the lower pair of signals $\phi_{12}=\phi_2+\pi/2-\phi_1$ and $\phi_{34}=\phi_4+\pi/2-\phi_3$ \footnote{{\color{black}Note that there is a $\pi/2$ phase difference} due to the $i$ factor in Eq. 1, and the maximum output intensity is achieved when sources 1 and 2 have $-\pi/2$ phase difference.}, the upper (H) and lower (V) interferometer have output signals with intensities:

\begin{equation}
	\begin{aligned}
		\mu_H = \mu_1/2 + \mu_2/2 + \cos(\phi_{12}) \sqrt{\mu_1\mu_2} ,\\
		\mu_V = \mu_3/2 + \mu_4/2 + \cos(\phi_{34}) \sqrt{\mu_3\mu_4} .\\
	\end{aligned}
\end{equation}

Taking into account that the H and V signals have randomized and independent global phases, we can simply consider them as two coherent states with random phases $\phi_H,\phi_V$ and intensities $\mu_H,\mu_V$. 



When these signals are combined at the PBS, the output state is still a phase-randomized coherent state with intensity $\mu=\mu_H+\mu_V$, and with each photon number states

\begin{equation}\label{eq:photon_number}
	\begin{aligned}
		\ket{n}_{\theta_{HV},\phi_{HV}} &= {1\over{\sqrt{n!}}}{a^{\dagger}}_{\theta_{HV},\phi_{HV}}^n \ket{vac}.
	\end{aligned}
\end{equation}

Here, the creation operator $a_{\theta_{HV},\phi_{HV}}^\dagger$ is defined by

\begin{equation}\label{eq:creation_operator}
	\begin{aligned}
		a_{\theta_{HV},\phi_{HV}}^\dagger &= \sqrt{\mu_H\over {\mu_H+\mu_V}} a^\dagger_H + e^{i(\phi_V - \phi_H)} \sqrt{\mu_V\over {\mu_H+\mu_V}} a^\dagger_V \\
		&= \cos{\theta_{HV}\over 2} a_H^\dagger + e^{i\phi_{HV}} \sin{\theta_{HV}\over 2} a_V^\dagger,\\
	\end{aligned}
\end{equation}

\noindent where we have defined $\theta_{HV}$ and $\phi_{HV}$ (the polar and azimuthal angles that uniquely determine a state on the Bloch sphere) as

\begin{equation}\label{eq:polarization}
	\begin{aligned}
		\theta_{HV} &= 2\cos^{-1} \left(\sqrt{\mu_H\over {\mu_H+\mu_V}}\right) ,\\
		\phi_{HV} &= \phi_V - \phi_H.
	\end{aligned}
\end{equation}

\begin{figure}[h]
	\includegraphics[scale=0.27]{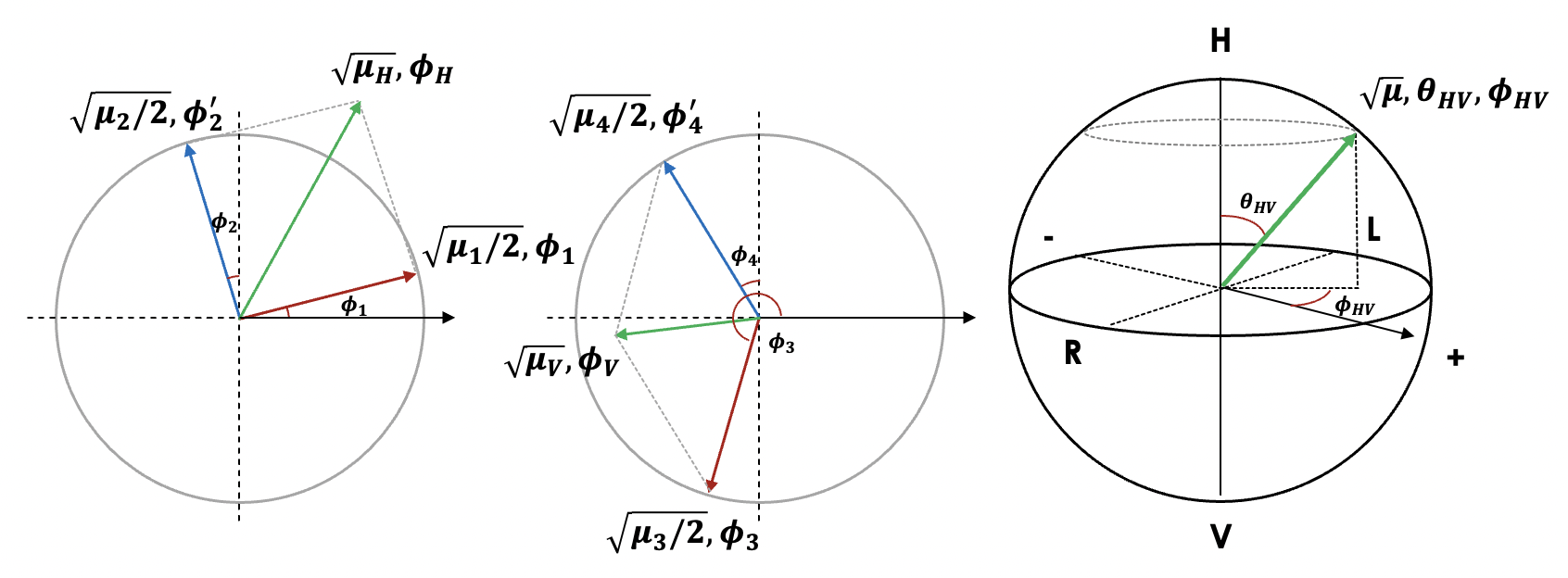}
	\caption{Conversion from the phases of the source signals $\{\phi_1,\phi_2,\phi_3,\phi_4\}$ to H and V signals $\{\mu_H,\phi_H,\mu_V,\phi_V\}$ and then to the final output state $\{\mu,\theta_{HV},\phi_{HV},\phi_G\}$. (Here we define $\phi_2'=\phi+\pi/2$ and $\phi_4'=\phi_4+\pi/2$, which are also uniformly random variables if $\phi_2,\phi_4$ are uniformly random.) We plot states as vectors on the complex plane, where the interference of each pair of signals can be viewed as a vector addition. The H and V output signals with $\mu_H,\phi_H$ and $\mu_V,\phi_V$ are combined at the polarizing beam splitter into the final output state, whose intensity is $\mu=\mu_H+\mu_V$, global phase is $\phi_G=\phi_H$, and the single-photon state on the Bloch sphere is described as a polar angle corresponding to the ratio of H and V components $\theta_{HV }=2\cos^{-1}(\sqrt{\mu_H/\mu})$ and the azimuthal angle corresponding to the phase difference between the H and V components $\phi_{HV}=\phi_V-\phi_H$. Note that, rigorously speaking, the output state is a coherent state, so the Bloch sphere only describes the state each photon is in. In other words, it describes the polarization mode the creation operator for the coherent state is in.}
	\label{fig:state}
\end{figure} 

The final output state of the source is a coherent state that can be described by the intensity $\mu$, the polar and azimuthal angles $\theta_{HV}$ and $\phi_{HV}$ that determine the polarization, and a global random phase $\phi_G=\phi_H$. If the input phases $\{\phi_1,\phi_2,\phi_3,\phi_4\}$ are uniformly random, and if we set $\mu_1=\mu_2=\mu_3=\mu_4=\mu_{max}/2$, we are in fact able to generate any output state with arbitrary intensity $0 \leq \mu \leq \mu_{max}$ \footnote{In fact, the maximum intensity the source can output is $\mu_1+\mu_2+\mu_3+\mu_4=2\mu_{max}$, but only with a single fixed polarization of $\theta_{HV}=2cos^{-1}(\mu_{max}/2\mu_{max})=\pi/2$. A state with output intensity $0 \leq \mu \leq \mu_{max}$ can take any arbitrary polarization angle, while the range of possible polarization angles decreases as $\mu$ exceeds $\mu_{max}$ and approaches maximum intensity.} and with arbitrary polarization. A more detailed proof is included in Appendix A. Again, note that here the output state is determined by the four degrees-of-freedom that come from the four independent phases of the sources $\{\phi_1,\phi_2,\phi_3,\phi_4\}$, which are first mapped to $\{\mu_H,\mu_V,\phi_H,\phi_V\}$ and then to $\{\mu,\theta_{HV},\phi_{HV},\phi_{G}\}$. An illustration of such a mapping is shown in Fig. \ref{fig:state}, and a mathematical description of the mapping process is also included in Appendix A.

So far, we have proposed a fully passive source capable of generating a phase-randomized coherent state with arbitrary intensity and polarization. Note that such a source is protocol-independent, and can in principle be used to prepare different states for various QKD protocols. The actual states Alice prepares will be determined by how she post-selects the intensity and polarization on the output state, depending on her local measurement results. In the next subsection we will describe example strategies that Alice can perform, to e.g. implement a source for BB84 protocol.

Though we have focused on polarization encoding in the above formulation, our encoding scheme can work as long as there are two signals that come from independent modes and have arbitrary intensities (created from the first step of passive decoy post-selection). The post-selection on the intensities and the Bloch sphere representation of the post-selected regions would be the same regardless of what the two independent modes actually are. For instance, the H and V modes can be replaced with early and late time bins respectively, {\color{black} while the PBS would be replaced by a BS}. In this case the Z basis would be time-bin encoding, and the X/Y bases would be in phase-encoding. In fact, such a time-bin phase encoding source would likely be ideal to implement a fully passive RFI-QKD scheme in a fibre-based system, since the time-bin basis is relatively stable and the phase basis would drift over time, which can be compensated by an RFI scheme.

\section{Protocol}

As we described in the previous section, our fully passive source can output a state with arbitrary intensity and polarization, making it general-purpose and protocol-independent. In practice, Alice needs to decide on a set of post-selection strategies to determine the states she intends to prepare, depending on the QKD protocol she chooses.

Here as an example, we show how she can prepare six polarization states $\{H,V,+,-,L,R\}$ in three bases. This set of states (or a subset of it) can be used in various protocols such as BB84, RFI-QKD, and the six-state protocol. Note that if Alice and Bob each hold such a source and perform such a post-selection, they can also use it for MDI-QKD (with X and Z bases) or RFI-MDI-QKD (each with all three bases) to implement a protocol without side-channels in either modulators or detectors. Again, we stress that what we show here is an example of a set of strategies Alice can choose, but she can also choose other strategies to prepare any polarization state or decoy intensities.\\

This section is structured as follows:

\begin{enumerate}
	\item In Section III.A we present as an example one possible set of post-selection regions Alice can use to define the states she prepares;
	\item In Section III.B we discuss an important technique Alice can use to arbitrarily shape the probability distribution for her signals, which not only can be used for correcting non-uniform phase distributions, but also plays a key role with enabling decoy-state analysis for passive sources (as we discuss in Section III.D);
	\item In Section III.C we explain that mixed polarizations in the single photon component of the signal state do not affect security (but only add to the observed noise);
	\item In Section III.D we propose a technique to make the observations from the decoy-state signals (which are phase-randomized coherent states with both mixed intensities and mixed polarizations) compatible with the decoy-state analysis, hence overcoming one of the main obstacles of fully passive QKD, the correlations between intensity and polarization;
	\item In Section III.E we discuss some practical considerations for experimentalists when implementing the system.
\end{enumerate}   

\subsection{State Preparation}

Alice can observe and post-select based on three variables she observes, $\{\mu_H,\mu_V,\phi_{HV}\}$, which correspond to the outcomes of her two intensity measurements using classical photodiodes and her polarization measurement. These three variables relate one-to-one to the output state $\{\mu,\theta_{HV},\phi_{HV}\}$ based on $\mu=\mu_H+\mu_V$ and $\theta_{HV}=2\cos^{-1}(\sqrt{\mu_H/(\mu_H+\mu_V)})$. Note that the global phase $\phi_G$ (the fourth degree-of-freedom) is not post-selected, since it needs to be randomized and also unknown to Eve. This allows us to employ the photon-number picture and hence apply the decoy analysis to the output states.

\begin{figure}[h]
	\includegraphics[scale=0.255]{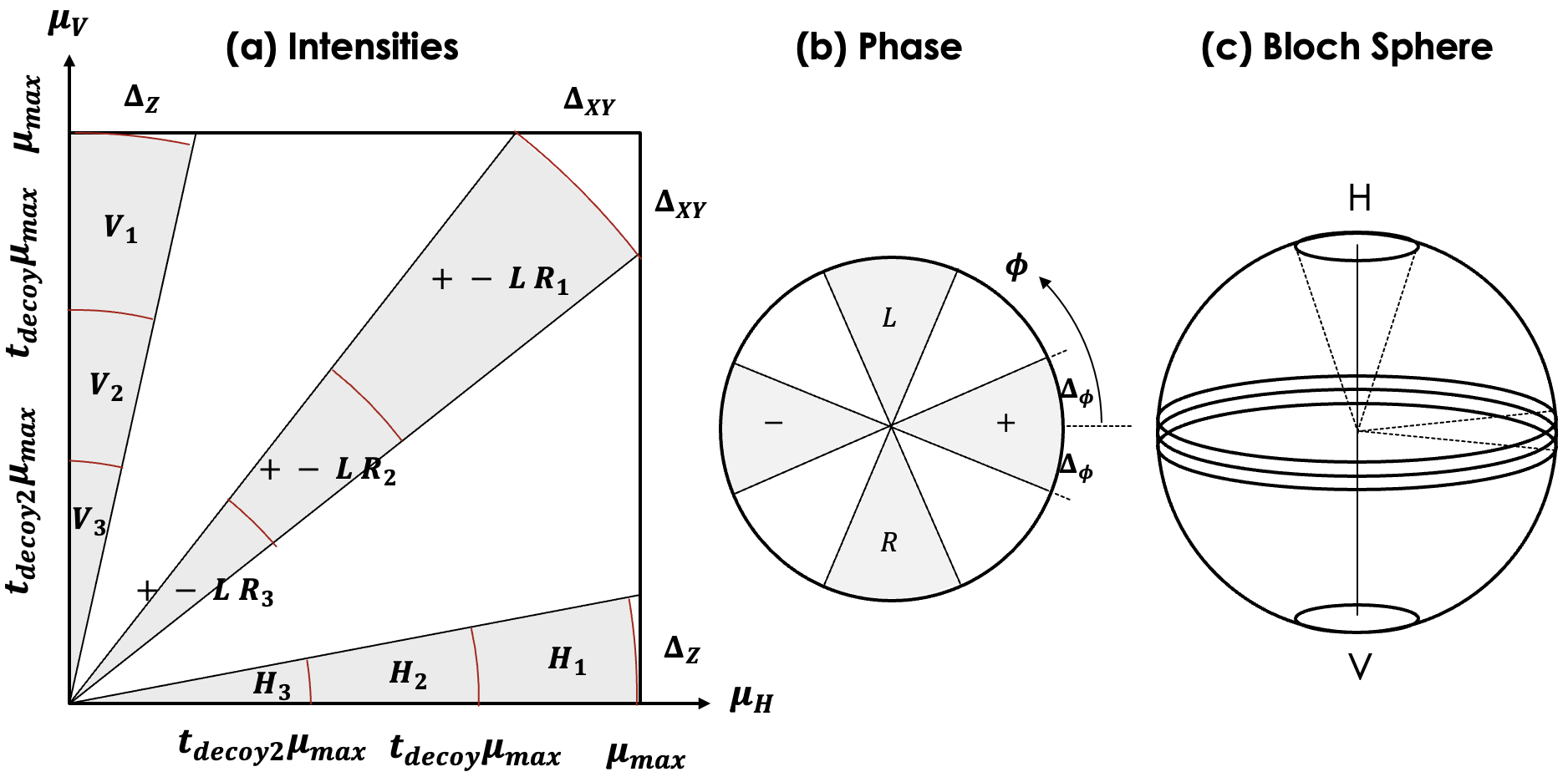}
	\caption{(a) Post-selection on intensities $\mu_H,\mu_V$. The highlighted regions correspond to the Z basis and X-Y plane on the Bloch sphere. Note that the slope $x=\mu_V/\mu_H$ on this plot determines the latitude (polar angle) $\theta_{HV}$ of the state on the Bloch sphere, i.e. the proportion of H versus V state components. The post-selected regions, e.g. the X-Y signals here, can further be divided into sub-regions to implement decoy states. (b) The additional polarization (phase) post-selection step that determines the longitude of the state. We can utilize this step to create e.g. X basis and Y basis polarization states, which are shown as highlighted regions in the figure. (c) Illustration of the corresponding states on the Bloch sphere based on the intensity post-selection step.}
	\label{fig:regions}
\end{figure}

To prepare the states $\{H,V,+,-,L,R\}$, Alice can define regions in the $\{\mu_H,\mu_V,\phi_{HV}\}$ space, such as shown in Fig. \ref{fig:regions} (a) and (b) corresponding to the set of desired states. In addition, she can divide the regions into sub-regions that have similar polarization but different intensity distribution, which allows her to perform the decoy-state analysis. The post-selection on $\mu_H,\mu_V$ and $\phi_{HV}$ are all independent, which allows us to simultaneously perform passive encoding and passive decoy-state choice, i.e. to implement a fully passive QKD source. Note that, while we divide the $(\mu_H,\mu_V)$ space according to the polarization $\theta_{HV}$, each of these areas are further bound by a maximum radius and form \textit{concentric sector shapes}. This choice of post-selection region in fact is crucial for enabling the decoy-state analysis, which will be further explained in a later subsection.

The size of these post-selection regions determines the sifting probability and the inherent error rate. Finer regions result in a distribution closer to the desired polarization state, but demand more signals to be discarded, which lowers sifting probability. This means that the region size must be subject to optimization. The source preparation and post-selection regions can be fully described by six parameters:

\begin{equation}
	[\mu_{max},\Delta_Z,\Delta_{XY},\Delta_\phi,t_{decoy},t_{decoy2}]
\end{equation}

\noindent where the maximum intensity $\mu_{max}$ on the H or V mode (which is twice the intensity of each of the four laser sources, assuming all the four sources have equal intensity) defines be boundaries of the post-selection. The other parameters are the post-selection slices on polarization angle $\Delta_Z$ for the Z basis and $\Delta_{XY}$ for the X/Y basis, the post-selection slice $\Delta_\phi$ on the relative phase between the H and V components, and the two decoy settings (which scale down the post-selection region by multiplying the sector radius by the factors of $t_{decoy}$ and $t_{decoy2}$, respectively). 

Importantly, unlike active QKD where the decoy settings are discrete intensity levels, here the decoy settings are defined by continuous post-selection regions (which we denote by $S_i$). Also, note that, the decoy settings can \textit{overlap} with each other (in other words, a region can be a subset of another region. For instance, we can directly use three sectors, each larger than the one before it, instead of the non-overlapping regions between the sectors, as the decoy regions), since the decoy-state analysis solves linear equations, which we can freely combine with or subtract from one another to form new linear constraints. \footnote{In fact, choosing overlapping regions might be slightly advantageous in the finite-size scenario since data for some decoy settings are combined.}

These parameters define all the six possible polarization states on the three bases X, Y, and Z, each having three decoy settings (i.e. a total of 18 possible usable states). Of course, if the BB84 protocol is implemented, Alice chooses only 12 of those states from two bases.

Having defined the post-selection regions, the expected value of any observable $Q$ (which can be deterministically calculated for any given set of $\mu_H,\mu_V,\phi_{HV}$) is an integration over the post-selection region $S_i$

\begin{equation}
	\begin{aligned}
		\langle Q \rangle &= (1/P_{S_i}) \iiint _{S_i} p(\mu_H,\mu_V,\phi_{HV})\\
		&\times Q(\mu_H,\mu_V,\phi_{HV}) d\mu_H d\mu_V d\phi_{HV},\\
		P_{S_i} &= \iiint _{S_i} p(\mu_H,\mu_V,\phi_{HV})d\mu_H d\mu_V d\phi_{HV},
	\end{aligned}
\end{equation}

All the decoy-states and encoding states are simply different integration regions for the variables. $p(\mu_H,\mu_V,\phi_{HV})=p_1(\mu_H,\mu_V)p_2(\phi_{HV})$ is the classical probability of obtaining a given combination of variables, which is a uniform distribution for $\phi_{HV}$, while the distribution for $\mu_H$ and $\mu_V$ can be obtained from the uniform distributions of $\phi_{12}$ and $\phi_{34}$ by using the bivariate transformation theorem:

\begin{equation}
	\begin{aligned}
		p_{\mu}(\mu_H,\mu_V) &= {1 \over {\pi^2 \sqrt{\mu_H (\mu_{max} - \mu_H)\mu_V (\mu_{max} - \mu_V)}}}, \\
		p_{\phi}(\phi_{HV}) &= {1\over {2\pi}}.
	\end{aligned}
\end{equation}

More details on the channel model and the simulation of the observables can be found in Appendix \ref{ref:appendix_channel}.


\subsection{Post-selection on the Probability Distribution}

In this subsection, we propose a very useful tool for Alice that allows her to arbitrarily shape the intensity and phase probability distributions $p(\mu_H,\mu_V,\phi_{HV})$ of her source. This technique not only greatly increases the amount of freedom Alice has in deciding the output states of her source (in a way playing the role of a modulator, but without introducing active modulation or side-channels), but also plays a crucial role in guaranteeing the security of passive QKD, as we show in the following subsections.

Note that, in the state preparation process described so far in the previous subsection, Alice only passively observes $\{\mu_H,\mu_V,\phi_{HV}\}$ (which is sampled from a fixed distribution $p(\mu_H,\mu_V,\phi_{HV})$ determined by her four laser sources) and defines regions to post-select them. Here we propose that, on top of the post-selection regions, Alice can also perform an ``additional post-selection technique" where she keeps or throws away each signal according to an arbitrary probability distribution $q(\mu_H,\mu_V,\phi_{HV})$ that she chooses. This allows Alice to \textit{arbitrarily shape} the intensity and phase probability distributions at her will, resulting in a final probability distribution $pq$ (subject to normalization).

\begin{figure}[h]
	\includegraphics[scale=0.3]{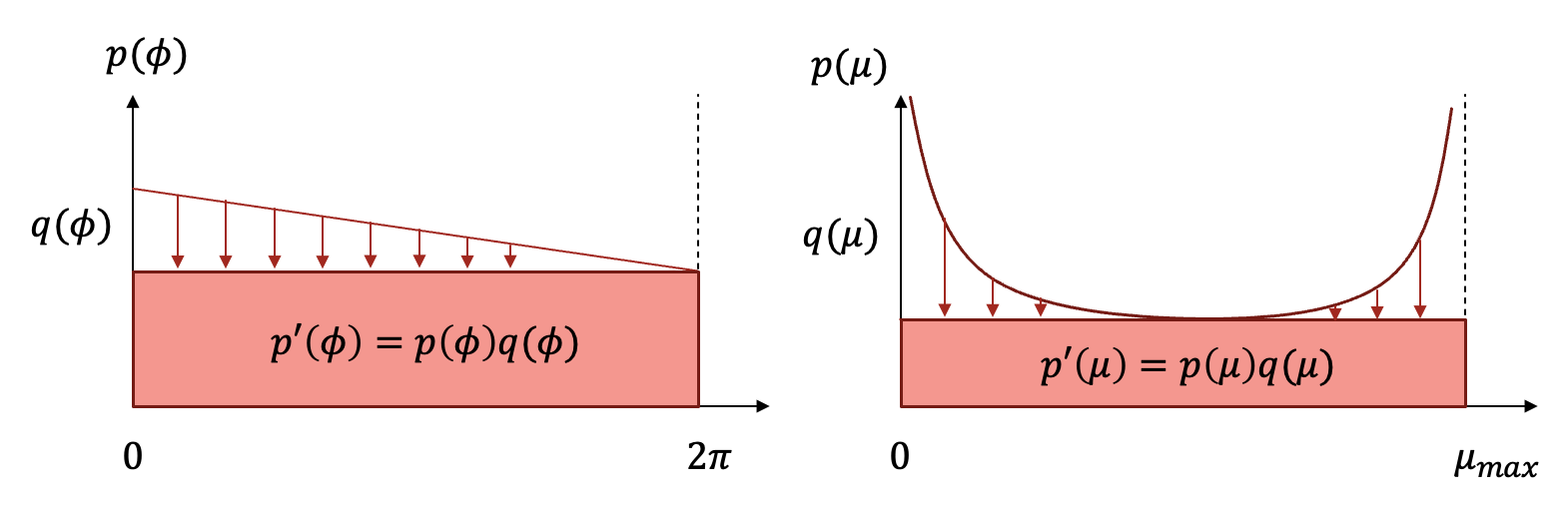}
	\caption{Examples of additional post-selection by Alice. Here if Alice can characterize a known distribution, e.g. $p(\phi)$ for the phase, even if the distribution is not uniform, she can apply an additional custom post-selection procedure to selectively discard/keep signals randomly with a passing probability $q(\mu)$. The resulting distribution $p'(\phi)=p(\phi)q(\phi)$ (after normalization by $\int_0^{2\pi} q(\phi) d\phi$) can be arbitrarily shaped based on $q(\phi)$. For instance, we can shape a known but non-uniform distribution into a uniform distribution. Similar conclusion applies to the intensity distribution $p(\mu)$, which, again, Alice can shape into a distribution she desires, such as a uniform distribution or any other compatible distribution.}
	\label{fig:postselection}
\end{figure}

The first use case for such a post-selection is if a phase distribution $p_\phi(\phi)$ is non-uniform, but follows a known distribution (here $\phi$ could be any of the phases $\phi_1,\phi_2,\phi_3,\phi_4$). Alice can now apply an additional post-selection to discard part of the signals and shape the distribution into a uniform one. This is particularly important because in passive QKD we cannot use a phase modulator to uniformly and randomly modulate the phase of each pulse. Instead, the security is subject to the inherent uniformness of the phase distribution of the laser sources. Using such additional post-selection technique, the above restriction can be loosened, since we can always characterize and shape the distribution to uniform at the expense of discarding some signals, as shown in Fig. \ref{fig:postselection} left plot.

Another use case is shaping the probability distribution of the intensities $\mu_H$ and $\mu_V$, as shown in Fig. \ref{fig:postselection} right plot. This will play a crucial role in enabling the decoy-state analysis for passive QKD, because in the passive source, polarization and intensity are \textit{coupled}, meaning that we usually cannot construct a linear program for the decoy-state analysis. Using the additional post-selection technique, she can arbitrarily design a probability distribution that is decoupled in terms of polarization angle and intensity, thus enabling the decoy-state analysis. We will later explain this in more detail in Sec. III. D.

The key message is, as long as there is some source of randomness (such as the phase randomness of the sources) that Alice can use as a starting-point, Alice can -- in a way \textit{actively} -- modulate the probability distribution arbitrarily, according to some local random bits she holds (which could come from e.g. a local random number generator). She then announces the status of successful/failed post-selection of each event to Bob. 

Importantly, such a process does not introduce side-channels like physical modulators, since this process happens as part of the post-selection too, inside Alice's classical post-processing device. The price Alice pays, though, is the need to use locally generated random bits, as well as discarding some signals (which has little effect on the key rate for asymptotic case, but can decrease the performance in the finite-size regime). However, as we show in Sec. IV, the passive BB84 protocol for instance still has relatively good key rate even with extensive post-selection on the intensity distribution.

\subsection{Security of Mixed State Sources}

A main difference of passive QKD is that the source is post-selected. This means that, as long as the selection regions are finite, the signals being sent out would actually be a mixture of various intensities and polarizations. The biggest problem one needs to consider is whether this might affect the security of the protocol. 

There are two steps in showing the security of passive QKD: 

\begin{enumerate}
	\item Using the observed statistics from phase-randomized coherent states (including all decoy settings), we show that we can use the decoy-state analysis to reliably obtain the upper and lower bounds for the yield and QBER of \textit{perfectly encoded single-photons}, even if the coherent states are prepared with mixed (and even coupled) polarizations and intensities \footnote{If needed, using our decoy-state analysis we can additionally also obtain statistics (such as QBER) for single photons with mixed polarizations.};
	\item Our setup is equivalent to a \textit{single-photon source} that emits photons with mixed polarizations in the signal basis and ``perfect" (i.e. in ideal pure states) polarizations in the test bases. Such state preparation imperfection in the signal basis can be viewed of as \textit{classical noise} in Alice's post-processing, which does not increase the amount of privacy amplification that one needs to apply, meaning that we can calculate the privacy amplification amount using bounds on perfect single photons statistics obtained from our decoy-state analysis.
\end{enumerate}

We will discuss the decoy-state analysis in point (1) in the next subsection. Throughout this subsection, we will focus on point (2) and limit our discussion to only the single-photon components, i.e. we will consider a single-photon source that emits mixed states in a given basis. 

\begin{figure}[h]
	\includegraphics[scale=0.4]{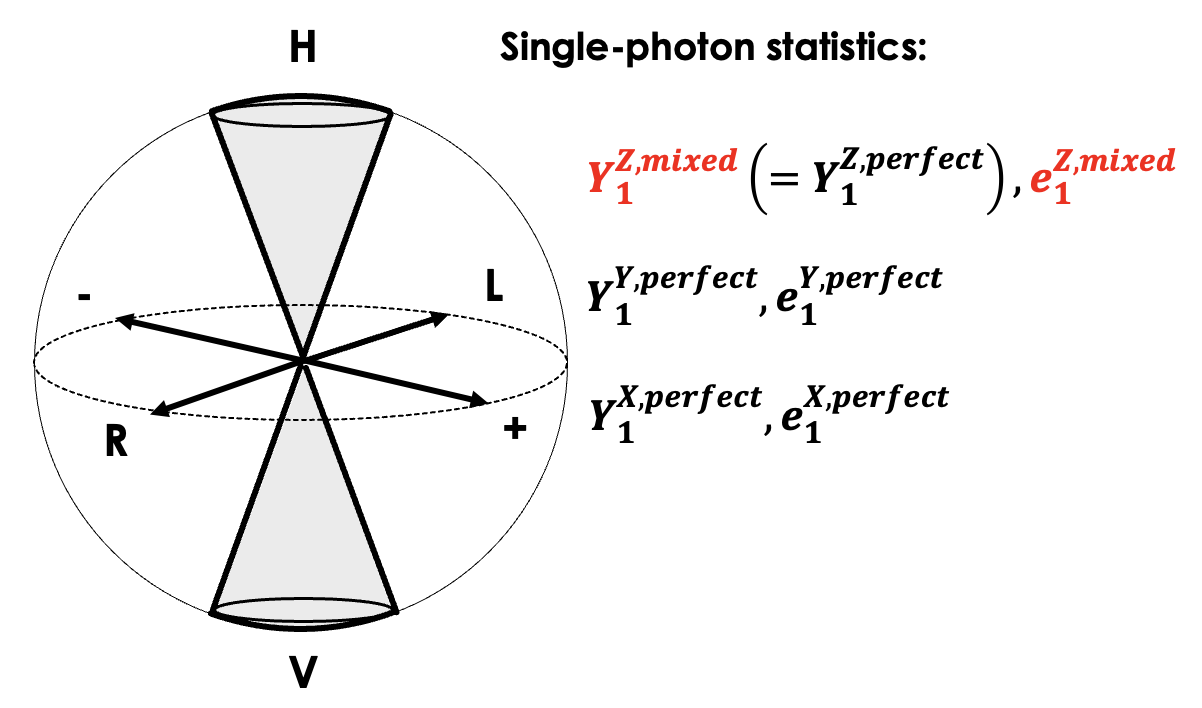}
	\caption{An illustration for a single-photon-based protocol with a source that emits perfectly encoded single photons in X and Y bases, but single photons with mixed polarizations in the Z basis. Our goal is to bound the security key rate for such a protocol, using the single-photon statistics obtained from decoy-state analysis. Note that from the decoy state analysis it is possible to obtain bounds on the statistics for both perfectly-encoded photons and, if necessary, single photons with mixed polarizations, such as $e_1^{Z,mixed}$ (which is not used for e.g. BB84, but can play a role in protocols that use Devetak-Winter bound \cite{DevetakWinter}, such as RFI-QKD \cite{RFIQKD}). Also, $Y_1^{Z,mixed}$ is equal to the yield of perfectly encoded single photons in Z basis (as long as the H and V density matrices still add up to identity).}
	\label{fig:signal}
\end{figure}

Here we consider protocols where the Z basis is used for key generation. Similar arguments in this subsection would apply to cases where the X or Y bases are used for key generation. Since we already apply the decoy-state analysis on the test states (which are coherent states with mixed polarizations and intensities) to obtain bounds on single photons, we can consider an equivalent ``virtual protocol" where Alice uses a single-photon source that emits \textit{perfectly encoded} single photons in the X and Y bases as test states and single photons with \textit{mixed polarizations} only in the Z basis as signal states, as illustrated in Fig. \ref{fig:signal}. Importantly, in this step we no longer need to consider imperfectly encoded single-photon test states in the X and Y bases anymore, because the decoy-state analysis in the first step already takes care of the imperfectly encoded coherent states and lets us obtain the bounds on perfectly encoded single-photon statistics (in all of X, Y, Z bases). \footnote{The bounds for the Z basis are slightly different: since the single-photon yield is basis-independent, the mixed polarization doesn't affect it, i.e. $Y_1^{Z,mixed}$ is simply $Y_1^{Z,perfect}$. Meanwhile, $e_1^{Z,mixed}$ can be obtained from our decoy-state analysis if needed. It is not used in BB84, but may appear in protocols, such as RFI-QKD \cite{RFIQKD}, that use the Devetak-Winter bound \cite{DevetakWinter}.} Therefore, the problem becomes: suppose we already have bounds on single-photon statistics in all bases (we can denote them as $\{Y_1^{X,perfect},Y_1^{Y,perfect},Y_1^{Z,mixed}\}$ and $\{e_1^{X,perfect},e_1^{Y,perfect}, e_1^{Z,mixed}\}$), how do we calculate the key rate of such a protocol with imperfect Z basis signal states?

Here, the intuition is that as long as the state being sent out still averages to a fully-mixed state for any basis, Eve is not able to gain any additional information on the key. When Alice prepares the Z basis signals, due to the finite-size of the post-selection region, the actual states are not be pure polarization states but in states $\rho_H$ and $\rho_V$, which are mixed states that depend on the definition of the corresponding post-selection regions. As long as the distributions of the polarization fluctuations in the mixture are symmetric, the encoded states always satisfy:

\begin{equation}\label{eq:fully_mixed}
	\begin{aligned}
		\rho_H + \rho_V = I\\
	\end{aligned}
\end{equation}

\noindent and therefore do not leak information to Eve about Alice's basis choice even if the signal states are mixed.

Next, we rigorously show that the security is not affected, by showing that \textit{the imperfect preparation of single photons in the Z basis is equivalent to trusted classical noise in Alice's post-processing}, meaning that the imperfect state preparation only increases the QBER but does not affect the amount of privacy amplification.

To illustrate this point, we consider an equivalent entanglement-distribution picture, where an entanglement source in Alice's lab sends perfect EPR pairs to Alice and Bob, who both performs measurements. In the ideal case with no state preparation flaws, Alice measures with POVMs:

\begin{equation}\label{eq:POVM}
	\begin{aligned}
		P_Z^0&=\ket{H}\bra{H}; \\
		P_Z^1&=\ket{V}\bra{V}; \\
		P_X^0&=\ket{+}\bra{+}; \\
		P_X^1&=\ket{-}\bra{-}; \\
		P_Y^0&=\ket{L}\bra{L}; \\
		P_Y^1&=\ket{R}\bra{R}; \\
	\end{aligned}
\end{equation}

In the non-ideal case in Fig. \ref{fig:signal}, Alice's imperfect preparation of Z basis states in the prepare-and-measure picture can be described here as \textit{imperfect measurements}. 


We observe the fact that the polarization fluctuation in the source is symmetric, i.e. the distribution of polarization angles is ``centered at" the perfectly prepared states. First, let us consider a simple two-dimensional rotation case (where there is no phase $\phi$ fluctuation, and the polarization fluctuations are simply misalignments on the X-Z plane). Suppose that the Z basis states $\ket{H},\ket{V}$ suffer from a random $\theta$ deviation from the originally desired polarization. In the prepare-and-measure picture, when Alice wants to prepare a pure state $\ket{H}$, due to the imperfect encoding, she actually prepares a state $\ket{H(\theta)}= \cos\theta \ket{H} + \sin\theta \ket{V}$. The angle $\theta$ could follow a distribution $p_\theta(\theta)$. Here, the key requirement is that

\begin{equation}
	p_\theta(-\theta)=p_\theta(\theta),
\end{equation}

\noindent i.e. $\langle \theta \rangle=0$, or that the distribution is ``centered at" the perfect case where $\theta=0$. Now, due to the symmetry of the distribution, we can always find a pair of misaligned states with equal probability:

\begin{equation}\label{eq:depolarize}
	\begin{aligned}
		\ket{H(\theta)} = \cos\theta \ket{H} + \sin\theta \ket{V}, \\
		\ket{H(-\theta)} = \cos\theta \ket{H} - \sin\theta \ket{V}, \\
	\end{aligned}
\end{equation}

In the entanglement distribution picture, instead of preparing imperfect states $\ket{H(\theta)}$ and $\ket{H(-\theta)}$, Alice could have measured with imperfect POVMs $\ket{H(\theta)}\bra{H(\theta)}$ and $\ket{H(-\theta)}\bra{H(-\theta)}$. The two POVMs are chosen with equal classical probabilities, which means that they sum up to the POVM:

\begin{equation}
	\begin{aligned}
		P_Z^{0'} &= {1\over 2} \left( \ket{H(\theta)}\bra{H(\theta)} + \ket{H(-\theta)}\bra{H(-\theta)} \right)\\
		&= (\cos^2\theta - \sin^2 \theta)\ket{H}\bra{H} + (\sin^2 \theta) I
	\end{aligned}
\end{equation}

\noindent which is a mixture between the perfect measurement $\ket{H}\bra{H}$ and a random measurement. The latter can be considered as a random \textit{noise} that is added to Alice's Z basis measurement results. It is equivalent to Alice simply adding random classical bit flips to her Z basis results with a probability of $sin^2(\theta)/2$ - which, of course, does not increase an eavesdropper's information.

In the actual passive source as shown in Fig. \ref{fig:signal}, the distributions of polarizations of prepares Z basis states are three-dimensional solid angles in the ``polar" regions of the Bloch sphere. In the entanglement distribution picture, Alice's Z basis measurements are integrated over these regions. When Alice desires to measure along $\ket{H}$ or $\ket{V}$, for instance, the actual measurement operators are:

\begin{equation}\label{eq:mixture}
	\begin{aligned}
		P_Z^{0'} &= \int_{0}^{\Delta_Z} \int_{0}^{2\pi} p(\theta_{HV},\phi_{HV}) \\
		&\times \ket{\psi'(\theta_{HV},\phi_{HV})} \bra{\psi'(\theta_{HV},\phi_{HV})} d\theta_{HV} d\phi_{HV}. \\
	    P_Z^{1'} &= \int_{\pi - \Delta_Z}^{\pi} \int_{0}^{2\pi} p(\theta_{HV},\phi_{HV}) \\
		&\times \ket{\psi'(\theta_{HV},\phi_{HV})} \bra{\psi'(\theta_{HV},\phi_{HV})} d\theta_{HV} d\phi_{HV}.
	\end{aligned}
\end{equation} 

\noindent where a given polarization state is defined by the angles $(\theta_{HV},\phi_{HV})$ (note that $\theta_{HV}$ is the polar angle on the Bloch sphere, which is equal to twice the polarization angle):

\begin{equation}
	\begin{aligned}
		 \ket{\psi'(\theta_{HV},\phi_{HV})} &= \cos{\theta_{HV}\over 2}\ket{H} + e^{i\phi_{HV}}\sin{\theta_{HV}\over 2}\ket{V}. \\
	\end{aligned}
\end{equation} 

The key point is that, as long as the distribution $p(\theta_{HV},\phi_{HV})$ satisfies 

\begin{equation}
	p(\theta_{HV},\phi_{HV})=p(\theta_{HV},\phi_{HV}+\pi),
\end{equation}

\noindent we can always divide the signals into symmetric pairs with polarizations $(\theta_{HV},\phi_{HV}),(\theta_{HV},\phi_{HV}+\pi)$, which, similar to our prior derivation, add up to a perfect state plus a fully-mixed state. This means that the entire integral in Eq. \ref{eq:mixture} would also add up to mixtures between the originally desired operator $\ket{H}\bra{H}$ or $\ket{V}\bra{V}$ and a random noise:

\begin{equation}
	\begin{aligned}
		P_Z^{0'} &= \lambda_Z \ket{H}\bra{H} + {1 \over 2} (1-\lambda_Z) I ,\\
		P_Z^{1'} &= \lambda_Z \ket{V}\bra{V} + {1 \over 2} (1-\lambda_Z) I ,\\
	\end{aligned}
\end{equation}

\noindent while the X and Y bases POVMs are the same as the perfect case in Eq. \ref{eq:POVM}.

Up to here, we have shown that the effect of imperfect preparation of Z basis signals is equivalent to classical post-processing noise in Alice's measurements, in the entanglement distribution picture. Such \textit{trusted noise} cannot possibly increase Eve's information, which means that it is okay to simply assume the same amount of privacy amplification as if Alice holds a perfect source. For instance, for BB84, using decoy-state analysis we can estimate $Y_1^{Z,perfect}$ (which equals $Y_1^{Z,mixed}$) and $e_1^{X,perfect}$, and it is still secure to use the privacy amplification $Y_1^{Z,perfect}[1-h_2(e_1^{X,perfect})]$ when calculating the final secure key rate. 

The implication is that the imperfect Z basis states have \textit{no adverse effect} on the privacy amplification, and the only penalties we receive from using a passive source are (1) increased Z basis QBER (hence more error-correction), (2) a more difficult decoy-state analysis to bound the perfect single-photon statistics using coherent states with mixed polarizations and intensities (which may result in looser bounds), and (3) more discarded signals due to sifting and post-selecting.

Another point worth noting is that, while here we choose to simply assume the same privacy amplification amount as a perfect source, such a process of adding noise may even decrease Eve's information on Alice's state \cite{AddingNoise}, and we can potentially apply the Devetak-Winter bound \cite{DevetakWinter} and incorporate such trusted noise to further improve the key rate. While a more rigorous study of applying such an approach to general passive QKD protocols will be the subject of our future studies, one simple example where the Devetak-Winter bound is used would be RFI-QKD \cite{RFIQKD}, for which the privacy amplification amount is $Y_1^{Z,mixed} \left( 1 - I_E\right)$, where

\begin{equation}
	I_E = (1-e_1^{Z,mixed}) h_2[(1+u_{\text{max}})/2)] + e_1^{Z,mixed} h_2[(1+v)/2)].
\end{equation}

\noindent where $u_{max}$ and $v$ are parameters related to the X and Y bases statistics and also $e_1^{Z,mixed}$(more details are included in Appendix \ref{ref:appendix_rate}). The important thing to note is that, unlike for the BB84 protocol above (where we used the Shor-Preskill key rate formula \cite{ShorPreskill} and the Z basis QBER does not affect privacy amplification), here $e_1^{Z,mixed}$ is used when estimating Eve's information. We show in the next subsection that such statistics for single photons with mixed polarizations can also be bounded from our decoy-state analysis too. Remarkably, one can see from simple calculations that an increased $e_1^{Z,mixed}$ due to mixed polarizations in Z basis signals actually serves to decrease Eve's information $I_E$ (and the amount of privacy amplification required).

\subsection{Decoy-State Analysis for Mixed State Sources}

The other problem we need to address is obtaining the bounds on single photon statistics using the data from test states, which are coherent states with mixed intensities and polarizations. There exists correlations between the intensities and polarizations of the states, which is the biggest obstacle to being able to apply the decoy-state analysis to the fully passive case, since the traditional decoy-state analysis does not work when estimating the yield $Y_1^{perfect}$ and the QBER $e_1^{perfect}$ of perfectly encoded single photons.

In this subsection we propose a method to estimate $Y_1^{perfect}$ and $e_1^{perfect}$ even when the source is mixed, by (1) carefully choosing a set of post-selection regions $S_i$ and (2) using post-selection to construct an intensity probability distribution $p_\mu$ that decouples the polarization angle and the intensity. This comes at a slight expense of additionally discarded data (which means a slight disadvantage in the finite-size regime, and no loss in the asymptotic limit, since, unlike signal states, the sifting factor for test states does not affect the key rate), but importantly enables the decoy analysis with our mixed source.\\

Let us first briefly review the standard decoy-state analysis \cite{decoystate_Hwang,decoystate_LMC,decoystate_Wang}. Given a set of intensities $\{\mu_i\}$, we can observe the gain and error-gain data $\{Q_{\mu_i}\}$ and $\{QE_{\mu_i}\}$. We can then establish a linear program to lower-bound the single-photon yield $Y_1$:

\begin{equation}
	\begin{aligned}
		Q_{\mu_i} = \sum_{n=0}^{\infty} P_n^{\mu_i} Y_n,
	\end{aligned}
\end{equation}

\noindent and a linear program to upper-bound the single-photon error-gain $e_1Y_1$:

\begin{equation}
	\begin{aligned}
		QE_{\mu_i} = \sum_{n=0}^{\infty} P_n^{\mu_i} e_n Y_n,
	\end{aligned}
\end{equation}

\noindent where $P_n^{\mu_i}=e^{-\mu_i}\mu_i^n/n!$ is the Poissonian distribution. Note that here $e_1Y_1$ is an individual variable in the linear program, and the single-photon QBER is upper-bounded by $e_1^U = e_1Y_1^U/Y_1^L$.\\

\begin{figure*}[t]
	\includegraphics[scale=0.565]{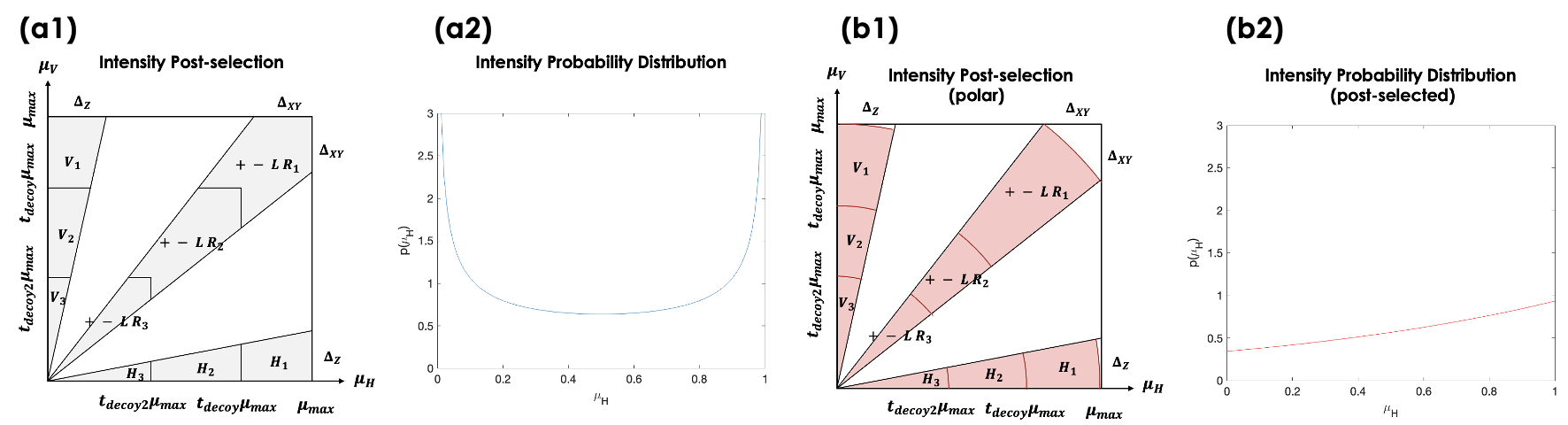}
	\caption{(a1) A set of "naive" post-selection regions determined by polarization angles and $max(\mu_H,\mu_V)$. (a2) Unmodified intensity probability distribution $p_{\mu_H}=1/(\pi\sqrt{\mu_H(\mu_{max}-\mu_H)})$. Same applies to $\mu_V$. (b1) The specially-designed sector shape post-selection regions. The decoy state setting is determined by the maximum radius, e.g. $\mu_{max},t_{decoy}\mu_{max}$ and $t_{decoy2}\mu_{max}$ for H state, where $t_{decoy}$ and $t_{decoy2}$ are scaling factors that can be optimized. (b2) The post-selected intensity probability distribution $p_{\mu_H} \propto exp(\mu_H)$ (where it is yet to be normalized and the plot shows the distribution after discarding part of the signals).}
	\label{fig:postselect_new}
\end{figure*} 

Now, for the passive QKD setup, the decoy states are no longer discrete intensity settings, but rather continuous regions $\{S_i\}$ that we use to perform post-selection. The linear constraints now become

\begin{equation}
	\begin{aligned}
		\langle Q \rangle_{S_i} = \sum_{n=0}^{\infty} \langle P_n Y_n \rangle_{S_i},\\
			\langle QE \rangle_{S_i} = \sum_{n=0}^{\infty} \langle P_n e_nY_n \rangle_{S_i},
	\end{aligned}
\end{equation}

\noindent where $P_n,Y_n$ and $e_nY_n$ are all functions of $(\mu_H,\mu_V)$, the intensities of the signals entering the PBS with H and V polarizations in the passive source \footnote{In fact, they are also functions of the relative phase $\phi_{HV}$, but we will leave that discussion for later in this subsection and focus on the two variables $(\mu_H,\mu_V)$ for the moment.}. Here $P_n(\mu_H,\mu_V)=exp(-\mu_H-\mu_V) (\mu_H+\mu_V)^n/n!$ is the Poissonian distribution. The expected value for any function of intensities $f(\mu_H,\mu_V)$ over a post-selection region $S_i$ is

\begin{equation}\label{eq:expectation}
	\begin{aligned}
		\langle f(\mu_H,\mu_V) \rangle_{S_i}&= {1\over P_{S_i}}\iint_{S_i} p_\mu(\mu_H,\mu_V)f(\mu_H,\mu_V) d\mu_H d\mu_V,\\
		P_{S_i} &= \iint_{S_i} p_\mu(\mu_H,\mu_V) d\mu_H d\mu_V,
	\end{aligned}
\end{equation}


\noindent where $P_{S_i}$ is simply the probability of choosing the given basis and decoy setting.\\

In this subsection, we make two key observations:

\begin{enumerate}
	\item For arbitrary choices of the post-selection regions (i.e. decoy settings) and the default intensity probability distribution $p_\mu$, a linear program generally cannot be constructed because $P_n$ and $Y_n$ both depend on the polarization and cannot be decoupled. 
	
	We prove this by showing that the ratio of the n-photon contributions $\langle P_nY_n\rangle_{S_i} / \langle P_nY_n \rangle_{S_j} $ for two different decoy regions $S_i$ and $S_j$ is not a known coefficient and is generally dependent on $Y_n$ (which is an unknown variable dependent on the channel), meaning that, except for special cases, $\langle P_nY_n\rangle_{S_i}$ cannot be written in the form of a coefficient specific to $S_i$ times a common variable such as $Y_n$ shared by all decoy settings. Therefore, a linear program cannot be constructed. Similar conclusion applies to $e_nY_n$.
	
	\item If we carefully choose a set of post-selection regions $S_i$ and construct specific probability distributions, it is still possible to construct a linear program and perform the decoy-state analysis.
	
	We will show that, with specific $S_i$ and $p_\mu$, we can define $\langle P_nY_n\rangle_{S_i} = \langle P_n\rangle_{S_i} Y_n'$, where $\langle P_n\rangle_{S_i} $ is only determined by the decoy setting $S_i$, and $Y_n'$ is a newly defined variable that is only determined by $Y_n$ and the polarization. This allows us to have a set of decoupled coefficients and variables for a linear program. Similar conclusion applies to $e_nY_n$.
\end{enumerate}

Here we make a key observation: the statistics $Y_n$ and $e_nY_n$ are not scalar values, but \textit{functions} of polarization angle $\theta=\cos^{-1}[\mu_H/(\mu_V+\mu_H)]$ \footnote{Note that here $\theta=\theta_{HV}/2$. As we will later explain, $\theta$ actually corresponds to the polar angle in the two-dimensional polar coordinates, which is more useful in the discussions here, while $\theta_{HV}$ is the polar angle on the Bloch sphere.}.  We can denote them as: \footnote{Here we have not yet specified the basis for the yield and error-yields. In each basis, the yield is usually the average of that of two states, for instance $Y_n^Z=(Y_n^H + Y_n^V)/ 2$. For convenience, we can simultaneously describe $Y_n^H$ and $Y_n^V$ with one angle $\theta$ by $Y_n^Z(\theta)= [Y_n(\theta) + Y_n(\pi/2-\theta)]/ 2$, so long as the distribution of polarization angles prepared by the passive source is symmetric with respect to H and V states. We will discuss this point in more detail in the end of this section.}

\begin{equation}
	\{Y_n(\theta),Y_ne_n(\theta)\}
\end{equation}

This actually leads to the biggest challenge for passive QKD: since the integral regions $S_i$ cover a range of possible $\theta$, this means that the variables $Y_n$ and $e_nY_n$ can no longer be decoupled from the integral, and by extension, they are coupled to the photon number distribution $P_n$. This is starkly different from active QKD, where the key assumption is that the variables $Y_n$ and $Y_ne_n$ are independent of the decoy settings, hence the photon number distributions $P_n$ are independent of $Y_n$ and $e_nY_n$, and we can construct a linear program and find the upper/lower bounds on these variables. 

The reason for such a dependency on polarization is explained below. First, it is rather easy to see that the error-yields $e_nY_n$ depend on the encoding polarization, since an imperfectly prepared polarization state results in a higher error rate. On the other hand, less straightforwardly, the yields $Y_n$ with $n\geq 2$ in fact might also depend on the polarization, since Eve may perform a photon-number-splitting attack and learn some information on the polarization by measuring the extra photons, and adjust the yield arbitrarily depending on the polarization. Note that, importantly, only $Y_1$ does not depend on imperfect polarization angles in the state preparation, so long as the imperfections (mixed polarizations) have symmetric distributions. For instance, $Y_1^Z=(Y_1^H+Y_1^V)/2$ will ideally stay the same even if the H and V states have mixed polarizations, as long as $\rho_H+\rho_V=I$. This is because Eve cannot obtain any information on the polarization (as we discussed in Sec. III. C). Similar conclusion applies to $Y_1^X$ and $Y_1^Y$.

For active QKD, the above dependency is not a problem, since ideally Alice always prepares each signal state (among $H,V,+,-,L$ and $R$) in exactly one perfect polarization, which uniquely determines a set of $P_n,Y_n$ and $e_nY_n$ for that particular signal state. On the other hand, for passive QKD, we have to integrate $P_n,Y_n$ and $e_nY_n$, which all depend on $\theta$, over an integral region $S_i$, so we cannot decouple them into a form such as $\langle P_n \rangle_{S_i} \langle Y_n \rangle_{S_i}$ (which is the form needed for a standard linear program). This breaks the key assumption for the decoy-state analysis and makes it no longer possible for us to straightforwardly construct a linear program. 

More rigorously speaking, for arbitrary unknown $Y_n$ and $e_nY_n$ (which are not scalar constants, but \textit{functions} of the polarization $\theta$), and for a set of any two post-selection regions $S_i$ and $S_j$, to be able to construct a linear program, any two n-photon terms $\langle P_nY_n\rangle_{S_i}$ and $\langle P_nY_n\rangle_{S_j}$ for these two decoy settings must satisfy $\langle P_nY_n\rangle_{S_i}/\langle P_nY_n\rangle_{S_j}=C_{S_i,S_j,n}$, where $C_{S_i,S_j,n}$ is a known constant coefficient that is independent of $Y_n$ and $e_nY_n$ (which corresponds to $P_n^{\mu_i}/P_n^{\mu_j}$ in active QKD). However, generally speaking, the above condition is \textit{not true} for any two selected regions $S_i$ and $S_j$. This means that, we generally cannot construct a set of consistent variables and known coefficients for them, making it not possible to set up a linear program. More detailed explanations and graphical illustrations of this problem are included in Appendix \ref{ref:appendix_decoy}.\\

To address this problem and construct a set of variables that are independent of the decoy setting, we choose (1) a set of concentric sector-shaped post-selection regions $S_i$, which range from $(\theta_{min},\theta_{max})$ and (2) a specially-constructed intensity probability distribution $p_\mu=C exp(\mu_H+\mu_V)$ (C is the normalization factor), using the aforementioned post-selection technique on the intensity. The new post-selection strategy (and a ``naive" strategy for comparison) are shown in Fig. \ref{fig:postselect_new}. Both the choice of $S_i$ and $p_\mu$ serve the purpose to decouple the polar and radial components of the integral in Eq. \ref{eq:expectation}.

To explain this more clearly, we can reformulate this problem in the polar coordinate, by converting the coordinate pair $(\mu_H,\mu_V)$ into $(r,\theta)$ instead where $r=\sqrt{\mu_H^2+\mu_V^2}$ and $\theta=tan^{-1}(\mu_V/\mu_H)$, integrated over the region $S_i$. (We will take the same approach for $e_nY_n$ and $Y_n$, so here we will focus on the yields first.) The integral looks like:

\begin{equation}\label{eq:expectation_polar}
	\begin{aligned}
		 \langle P_n Y_n \rangle_{S_i} &= {1\over P_{S_i}}\iint_{S_i}  p_\mu(r,\theta) P_n(r,\theta) Y_n(\theta) rdrd\theta, \\
	\end{aligned}
\end{equation}

\noindent where $P_n$ is the Poissonian distribution, and $p_\mu$ is the inherent intensity distribution from the source. Note that, importantly, $Y_n(\theta)$ is only a function of $\theta$ (i.e. the polarization) and not of $r$, since a Fock state does not keep any information of the original intensity of the pulse and Eve can at most obtain information on $\theta$. Same applies for the error-yield $e_nY_n$. 

Now, we choose $S_i$ as concentric sector-shaped regions as shown in Fig. \ref{fig:postselect_new} (b1), which in polar coordinates are simply rectangular regions, defined by $[\theta_{min},\theta_{max}]$ and $[0,r_{max,i}]$ (the only difference for different decoy settings would be the maximum radius, $r_{max,i}$). 

Furthermore, instead of using the original intensity distribution
\begin{equation}
	 p_\mu^{orig}=1/[{\pi^2 \sqrt{\mu_H(\mu_{max}-\mu_H)\mu_V(\mu_{max}-\mu_V)}}],
\end{equation}
\noindent from the passive source, we apply a step of additional post-selection on the intensity probability distribution, which randomly discards signals with probability $1-q_\mu$, i.e. the intensity probability distribution is multiplied by

\begin{equation}
	\begin{aligned}
		q_\mu = C\pi^2 \sqrt{(\mu_H(\mu_{max}-\mu_H)\mu_V(\mu_{max}-\mu_V))}e^{(\mu_H+\mu_V)},
	\end{aligned}
\end{equation}

\noindent such that the actual (conditional) intensity distribution $p_\mu$ becomes

\begin{equation}
	\begin{aligned}
		p_\mu=p_\mu^{orig}q_\mu = Ce^{(\mu_H+\mu_V)},
	\end{aligned}
\end{equation}

\noindent where $C$ is the normalization coefficient for the conditional distribution among the signals that passed the post-selection. \footnote{Note that if we are considering finite-size effects, though, instead of choosing a normalization coefficient, $C$ is a coefficient chosen such that all values of $q_\mu(\mu_H,\mu_V)\leq 1$ (since we are only allowed to \textit{discard} signals). Since the probability distributions for $(\mu_H,\mu_V)$ are decoupled, if we look at one input of the PBS (e.g. $\mu_H$) and the corresponding one-dimensional distribution, we should be looking for the largest $\sqrt{C}$ such that $\sqrt{C}exp(\mu_H)\leq 1/(\pi \sqrt{\mu_H(\mu_{max}-\mu_H)})$ over the entire region $[0,\mu_{max}]$. For instance, for $\mu_{max}=1$, $C\approx 0.12$. The total number of remaining signals after post-selection are $N'=Ce^{(\mu_H+\mu_V)}N$ for $N$ signals sent.} We chose this distribution since it cancels out with the exponential term $e^{-(\mu_H+\mu_V)}$ in the Poissonian distribution $P_n$. This allows the remaining function to be decoupled into $r$ and $\theta$ parts, although note that this may not be the only viable choice and other distributions might be able to decoupled radial and angular parts of Eq. \ref{eq:expectation_polar} too.

With the new $S_i$ and $p_\mu$ defined, we obtain

\begin{equation}
	\begin{aligned}
		&p_\mu(r,\theta) P_n(r,\theta) r \\
		=& C e^{r(\sin\theta+\cos\theta)}\\
		\times& e^{-r(\sin\theta+\cos\theta)}{{[r(\sin\theta+\cos\theta)]^n}\over n!} \times r \\
		=& C r^{n+1} (\sin\theta+\cos\theta)^n/n!, \\
	\end{aligned}
\end{equation}

\noindent hence we can simply rewrite Eq. \ref{eq:expectation_polar} as

\begin{equation}\label{eq:decouple_integral}
	\begin{aligned}
		\langle P_n Y_n \rangle_{S_i} &= {C\over P_{S_i}}\int_{0}^{r_{max,i}} (r^{n+1}/n!)dr\\
		&\times \int_{\theta_{min}}^{\theta_{max}} (\sin\theta+\cos\theta)^n Y_n(\theta) d\theta. \\
	\end{aligned}
\end{equation}

Note that, conveniently, the exponential term in $P_n$ cancels out with our specially constructed $p_\mu$, resulting in a probability distribution that can be decoupled into a direct product of polar and angular components.

We can further reshape Eq. \ref{eq:decouple_integral} by moving the constant coefficient of $\int_{\theta_{min}}^{\theta_{max}} (\sin\theta+\cos\theta)^n d\theta$ from the angular component to the radial one:

\begin{equation}
	\begin{aligned}
		&\langle P_n Y_n \rangle_{S_i} = {C\over P_{S_i}}\\
		&\times \left[\int_{0}^{r_{max,i}} (r^{n+1}/n!)dr\times {\int_{\theta_{min}}^{\theta_{max}} (\sin\theta+\cos\theta)^n d\theta}\right]\\
		&\times \left[{{\int_{\theta_{min}}^{\theta_{max}} (\sin\theta+\cos\theta)^n Y_n(\theta) d\theta} \over {\int_{\theta_{min}}^{\theta_{max}} (\sin\theta+\cos\theta)^n d\theta}}\right] \\
		&= {1\over P_{S_i}} \times \left[\int_{0}^{r_{max,i}} {\int_{\theta_{min}}^{\theta_{max}}  P_n(r,\theta)p_\mu(r,\theta)r dr d\theta}\right]\\
		&\times \left[{{\int_{\theta_{min}}^{\theta_{max}} (\sin\theta+\cos\theta)^n Y_n(\theta) d\theta} \over {\int_{\theta_{min}}^{\theta_{max}} (\sin\theta+\cos\theta)^n d\theta}}\right] \\
		&= \langle P_n \rangle_{S_i} \times Y_n',
	\end{aligned}
\end{equation}

\noindent where $\langle P_n \rangle_{S_i}$ can also be obtained using Eq. \ref{eq:expectation}, and we have defined $Y_n'$ (which we denote by ``pseudo-yield") as

\begin{equation}
	Y_n' = {{\int_{\theta_{min}}^{\theta_{max}} (\sin\theta+\cos\theta)^n Y_n(\theta) d\theta} \over {\int_{\theta_{min}}^{\theta_{max}} (\sin\theta+\cos\theta)^n d\theta}}.
\end{equation}

A similar argument applies to $e_nY_n'$ (``pseudo-error-yield") where we simply replace the $Y_n(\theta)$ above with $e_nY_n(\theta)$.

The key message here is that, by using the new $S_i$ and $p_\mu$, each n-photon term can be simply decoupled into $\langle P_n \rangle_{S_i}$ and a variable $Y_n'$ which is independent of the setting $S_i$, which can be used to construct a linear program 

\begin{equation}\label{eq:decoy}
	\begin{aligned}
		\langle Q \rangle_{S_i} = \sum_n \langle P_n \rangle_{S_i} \times Y_n' , \\
		\langle QE \rangle_{S_i} = \sum_n \langle P_n \rangle_{S_i} \times e_nY_n' .
	\end{aligned}
\end{equation}

Here, the pseudo-yield and pseudo-error-yield still nicely satisfy $Y_n',e_nY_n'\in[0,1]$, because $(\sin\theta+\cos\theta)^n = \sqrt{2}\sin^n(\theta+\pi/4) > 0$ over the region of $[0,\pi/2]$, and every single value of $Y_n(\theta)$ and $e_nY_n(\theta)$ belongs into the interval $[0,1]$. Therefore, from Eq. \ref{eq:decoy} we can use linear programming to find the upper and lower bounds on the pseudo-yields and pseudo-error-yields, $Y_n'^L$ and $e_nY_n'^U$.

As mentioned in Sec. III. C, the goal of the decoy state analysis is to obtain the upper and lower bounds on the yield and error-yield of perfectly prepared single photons. Firstly, we observe that, for single-photons, $Y_1$ does not depend on the polarization (since Eve cannot obtain any information on the polarization without disturbing the photon), hence the yield of perfectly prepared single photons is simply

\begin{equation}
	Y_1^{perfect} = Y_1' \geq Y_1'^L,
\end{equation}

\noindent which means that $Y_1'^L$ is also a lower bound for $Y_1^{perfect}$. We can therefore also denote it as $Y_1^{perfect,L}$.

On the other hand, for the error-yield, we can use the conclusion derived in the previous subsection. That is, the mixture of a pair of states respectively having a misalignment of $-\theta $ and $\theta$ from the ``perfect encoding" can be viewed as a pure state being depolarized, i.e. mixed with a fully mixed state (corresponding to a POVM mixed with random noise in the entanglement distribution picture). For instance, for the $\ket{H}$ state:

\begin{equation}
	\begin{aligned}
		\rho_H &=  (\cos^2\theta - \sin^2 \theta)\ket{H}\bra{H} + (\sin^2 \theta) I,
	\end{aligned}
\end{equation}

\noindent from which we can deduct that, since the QBER observed from the fully mixed state $I$ is always $50\%$, the average QBER for the mixed single photons (which is a weighted average of the perfectly encoded signals and the fully mixed state) cannot be smaller than the QBER of the perfectly prepared states:

\begin{equation}
	e_1Y_1^{perfect} \leq e_1Y_1(\theta).
\end{equation}

\noindent for every $\theta$. The equality sign is taken when $\theta$ corresponds to the perfectly prepared state ($\pi/4$ for the $+,-,L$ and $R$ states, $0$ for the $H$ state, and $\pi/2$ for the $V$ state). Note that this is applicable to the error-yield of a basis too, for instance $e_1^XY_1^X$ or $e_1^YY_1^Y$, and $e_1^ZY_1^Z$, where we can use one $\theta$ to represent a pair of states (which we will explain at the end of this subsection). This means that

\begin{equation}
	e_1Y_1^{perfect} \leq {{\int_{\theta_{min}}^{\theta_{max}} (\sin\theta+\cos\theta) e_1Y_1(\theta) d\theta} \over {\int_{\theta_{min}}^{\theta_{max}} (\sin\theta+\cos\theta) d\theta}} = e_1Y_1'.
\end{equation}

Therefore,

\begin{equation}
	\begin{aligned}
		e_1Y_1^{perfect} \leq e_1Y_1' \leq e_1Y_1'^U.
	\end{aligned}
\end{equation}

\noindent which means that the upper bound for the pseudo-error-yield is also an upper bound for the error-yield of ``perfectly encoded" single photons. Therefore, we can also denote $e_1Y_1'^U$ as $e_1Y_1^{perfect,U}$. One can simply divide $e_1Y_1^{perfect,U}$ by $Y_1^{perfect,L}$ to obtain the upper bound on the single-photon QBER $e_1^{perfect,U}$.\\

There are two additional points worth noting here. Firstly, for conciseness, so far we have only discussed the variables $(\mu_H,\mu_V)$, but in fact for the passive source, aside from the global phase randomization, there are three degrees-of-freedoms $(\mu_H,\mu_V,\phi_{HV})$, so $Y_n$ and $e_nY_n$ will also depend on $\phi_{HV}$, which is the relative phase between the $H$ and $V$ components. However, we note that the distribution for $\phi_{HV}$ is (1) ideally always a uniform distribution, and (2) always completely decoupled from the intensity distribution. Therefore, we can simply rewrite $Y_n(\theta)$ as

\begin{equation}\label{eq:wrapper}
	Y_n(\theta) = {{\int_{\phi_{HV,min}}^{\phi_{HV,max}}Y_n(\theta,\phi_{HV})p_\phi(\phi_{HV})d\phi_{HV}}\over {\int_{\phi_{HV,min}}^{\phi_{HV,max}}p_\phi(\phi_{HV})d\phi_{HV}}}
\end{equation}

\noindent where $p_\phi(\phi_{HV})$ is just a uniform distribution. Here we always first partially integrate over the phase variable $\phi_{HV}$, hence all the above expressions for $Y_n(\theta)$ can be kept unchanged.

Secondly, in the discussions above, we have deliberately omitted the actual state/bases prepared and simply described a general $Y_n(\theta)$ which is integrated over various regions. In fact, the above $Y_n(\theta)$ (where $\theta$ is simply the polarization angle $\theta=\theta_{HV}/2$, while $\theta_{HV}$ is the polar angle on the Bloch sphere) can be used to describe any of the six possible-to-prepare states $\{H,V,+,-,L,R\}$, depending on $\theta_{min},\theta_{max},\phi_{HV,min}$ and $\phi_{HV,max}$. 

For instance, the Z basis yield for active QKD is defined as:

\begin{equation}
	Y_n^Z = {Y_n^H + Y_n^V \over 2 },\\
\end{equation}

\noindent where ideally the H and V states are prepared with perfect polarization angles $0$ and $\pi/2$. Now for the passive source, instead of solving for the yields for the H and V states separately (which we could in principle do), for convenience we can pair up symmetric $H$ and $V$ signals (with polarization angles $0+\theta$ and $\pi/2 - \theta$) such that each single $\theta$ represents a pair of H and V signals. $Y_n^Z$ can be written as:

\begin{equation}
	Y_n^Z(\theta) = {{Y_n(\theta) + Y_n(\pi/2-\theta)} \over 2},
\end{equation}

\noindent where $\theta \in [0,\Delta_Z]$, and the Z basis states select all $\phi_{HV}\in[0,2\pi)$. Here if one integrates $Y_n^Z$ (or a function of $Y_n^Z$) over $[0,\Delta_Z]$, the polarization fluctuations in both the H and V signal states will be included.

On the other hand, the yields where Alice prepares X and Y bases can be written as

\begin{equation}
	\begin{aligned}
		Y_n^X(\theta) &= {1\over 2}\times {1\over{ 2\Delta_\phi}} [\int_{-\Delta_\phi}^{\Delta_\phi}Y_n(\theta,\phi_{HV})d\phi_{HV} \\
		&+ \int_{\pi-\Delta_\phi}^{\pi+\Delta_\phi} Y_n(\theta,\phi_{HV})d\phi_{HV}],\\
		Y_n^Y(\theta) &= {1\over 2}\times {1\over{ 2\Delta_\phi}} [\int_{\pi/2-\Delta_\phi}^{\pi/2+\Delta_\phi}Y_n(\theta,\phi_{HV})d\phi_{HV} \\
		&+ \int_{3\pi/2-\Delta_\phi}^{3\pi/2+\Delta_\phi} Y_n(\theta,\phi_{HV})d\phi_{HV}],\\
	\end{aligned}
\end{equation}

\noindent where $\theta \in [\pi/4-\Delta_{XY},\pi/4+\Delta_{XY}]$. Again, the symmetries of the signal states are utilized to define $Y_n^X$ and $Y_n^Y$ as functions of only a single variable $\theta$, each of which represents a pair of signals.

Overall, the key point is that $Y_n(\theta)$ is a general expression that can be used to describe statistics in different states (as long as the expression depends on $\theta$). The results in this whole subsection does not make any assumptions on $Y_n(\theta)$ (except that $Y_1$ does not depend on $\theta$), so the conclusions are not affected by the redefinitions of $Y_n(\theta)$ to account for different bases or to include phase distributions.\\

Up to here we have obtained the bounds for both the yield and QBER of perfectly prepared single-photon states, based on the original phase-randomized coherent states with mixed and coupled polarization and intensities. As we discussed in the previous subsection, these bounds on statistics $Y_1^{X,perfect,L},e_1^{X,perfect,U}$ and $Y_1^{Z,perfect,L}$ (Note that, since $Y_1$ does not depend on polarization, $Y_1^{Z,mixed}=Y_1^{Z,perfect}$) suffice in estimating the key rate of BB84, or other protocols that use Shor-Preskill key rate formula.

Additionally, if one makes use of Devetak-Winter bound, such as in RFI-QKD, the Z basis QBER for single photons with \textit{mixed} polarizations, $e_1^{Z,mixed,U}$, might be needed too. We make the simple observation here that 

\begin{equation}
	\begin{aligned}
		e_1Y_1^{Z,mixed} & =  \int_{0}^{\Delta_Z} p_\theta(\theta) \times e_1Y_1^Z(\theta) d\theta. \\
	\end{aligned}
\end{equation}

\noindent where we have omitted the variable $\phi_{HV}$ by first integrating over it, using Eq. \ref{eq:wrapper}. Note that $p(\theta_{HV})$ is simply the angular probability distribution, i.e. $p_\theta(\theta_{HV}) \propto (sin\theta+cos\theta)$. Therefore, given our sector-shaped post-selection regions $S_i$, we have

\begin{equation}
	\begin{aligned}
	e_1Y_1^{Z,mixed} &= {{\int_{\theta_{min}}^{\theta_{max}} (\sin\theta+\cos\theta) e_1Y_1^Z(\theta) d\theta} \over {\int_{\theta_{min}}^{\theta_{max}} (\sin\theta+\cos\theta) d\theta}}  &= e_1Y_1^{Z'}
	\end{aligned}
\end{equation}

Interestingly, this tells us that the pseudo-error-yields (as well as pseudo-yields) for single photons actually have physical meanings: they represent the yield and error-yield of single photons with \textit{mixed} polarizations due to passive encoding. Therefore, an upper (lower) bound $e_1Y_1^{Z',U}$ ($e_1Y_1^{Z',L}$) is also an upper (lower) bound for $e_1Y_1^{Z,mixed}$, meaning that we can also obtain the necessary statistics to calculate the Devetak-Winter key rate for e.g. RFI-QKD.

\subsection{Experimental Considerations}

Note that our proposed scheme depends on two key assumptions:

\begin{enumerate}
\item Randomness of phase: Each pulse from the source has a phase randomly and uniformly distributed between $[0,2\pi)$. Also, there is no phase correlation between neighbouring pulses or between any two sources.\\
\item Post-selection Accuracy: Alice can (1) accurately measure the intensities $\mu_H$ and $\mu_V$, and (2) accurately measure the polarization of the output state (i.e. she can measure the phase difference between H and V $\phi_{HV}$). Both measurements can be performed on strong light (which is attenuated to the single-photon level before sending it to the channel).\\
\end{enumerate}

Assumption 1 is in fact a fundamental prerequisite for active QKD too, where the security of the decoy-state analysis depends on the randomness of the phase. A main difference is that in an active setup, Alice is allowed to use an additional phase modulator which is driven by a true random source to actively randomize her phase, but for the passive setup, we need to depend on the inherent randomness coming from the sources. However, note that there have been successful implementations of QKD systems \cite{phaserandom1} that rely on the phase randomness of the source only, where gain-switched lasers are used and neighbouring pulses are considered independent and randomized. There are also works \cite{phaserandom2,phaserandom3} that studied the phase randomness of such sources and concluded that it is sufficient for use in QKD. Phase randomness in the laser source has also been successfully used for quantum random number generators \cite{QRNG1,QRNG2,QRNG3}.

Also, note that even if Alice's sources have a non-uniform but \textit{known} probability distribution $p(\phi)$ that she can accurately characterize, she can apply the additional post-selection method we presented in Sec. III.B, and ``shape" the distribution into a uniform one. This means that one can safely use these sources to construct the passive source.

Assumption 2 imposes requirements on Alice's measurement devices. The inaccuracy in her characterization of $\mu_H,\mu_V$ and $\phi_{HV}$ would mean looser bounds for post-selection, i.e. the actual distribution of the post-selected signals is slightly different from what we expect. As long as there is no bias between the inaccuracies in $\mu_H$ and $\mu_V$ (i.e. the output state is still fully mixed), there would not be any additional information leakage due to the imperfect encoding. However, the inaccuracy would affect the decoy-state analysis. Theoretically, its effect would be similar to \textit{intensity fluctuations}, which might cause Alice to incorrectly characterize her photon-number distribution and thus over-/underestimate the single-photon statistics. This may lead to an incorrect estimation of the amount of privacy amplification required. 

As long as Alice can characterize the amount of inaccuracy, though, Alice can incorporate it into her security analysis by e.g. taking the worst-case scenario given the fluctuations (which in the active case are fluctuations of the discrete intensity values, but here in the passive case would be fluctuations of the boundaries of the post-selection regions). Note that, since the signals in Alice's system are all at strong-light level (and are only attenuated to single-photon level right before being sent out), Alice can increase her intensity to decrease the relative inaccuracy caused by shot noise and other detector noises, albeit there is still a physical limitation to her maximum usable intensity (limited by her laser power and the requirement of linearity of her detectors). Additionally, her accuracy would be limited by the other detector noises (e.g. thermal noise) in her classical photodiodes. Such noise might be more significant for the passive setup, since here each pulse has a different intensity and one cannot integrate over multiple pulses to obtain an average intensity. Experimentalists might need to carefully choose the intensities, pulse width, and clock rate of their systems to minimize the inaccuracies caused by system noise.

The polarization/phase measurement (again, performed on classical light) could be implemented e.g. by using a tomography-like setup which splits the signal into two or three pairs of detectors (in $H,V$, $+,-,L$ and $R$ polarizations). Since Alice already knows $(\mu_H,\mu_V)$ from her intensity measurements, she only needs to determine $\phi_{HV}$. This means that she would only need two pairs of detectors (one pair is not sufficient since it cannot uniquely differentiate between e.g. $\phi_{HV}$ and $-\phi_{HV}$ with the same value $\cos\phi_{HV}$).

One more factor to consider is the quality of the interference in the source, which depends on how similar (in terms of e.g. frequency, timing, and pulse shape, such that the signals are on the same mode) can the four independent sources generate the signal pulses. Reliable interference from commercial off-the-shelf independent lasers have been reported, e.g. in studies on interference between WCP sources \cite{interference1,interference2}, in the implementation of MDI-QKD protocols, such as Refs. \cite{MDIQKDexp1,MDIQKDexp2} (remarkably, Ref. \cite{MDIQKDexp2} did not even use active phase randomization and only relies on random phases between pulses from gain-switched lasers that optically seed secondary lasers), and in CV-QKD with a true local oscillator laser \cite{CVQKD_LO}. In addition to interference visibility, the mismatch in laser frequencies would also cause the problem of phase drift in the H and V pulses. Experimentalists need to select lasers with stabilized central frequencies, and having a moderate coherence time such that the phase is stable within each pulse but random from pulse to pulse (one can even e.g. select every other pulse or every few pulses to guarantee phase randomness). One can also use gain-switched lasers for phase randomization, such that they will not be limited by the coherence time of the lasers. The pulse width may also need to be reduced in order to mitigate the effect of phase drift within a pulse. On the other hand, the minimum pulse width is limited by the sample rate, resolution, as well as by the noise in Alice's classical detectors used for intensity measurements, so a trade-off is present in the choice of pulse width between phase stability and feasibility/accuracy of intensity detection.

\begin{figure}[h]
	\includegraphics[scale=0.275]{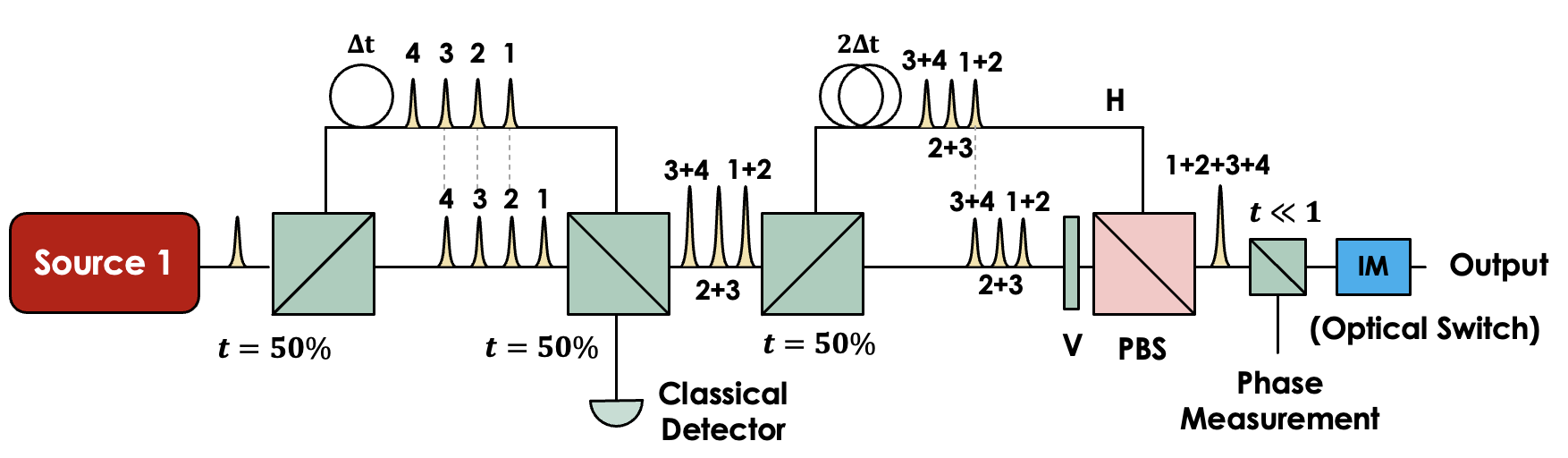}
	\caption{An equivalent setup for a fully passive source using only one single source. This setup makes use of two delay lines of $\Delta t$ and $2\Delta t$ to interfere four independent pulses from different time bins. Suppose we index the pulses in time order as pulses 1, 2, 3, 4, etc.. The first interferometer functions like the passive decoy part respectively for the H and V arms, with pulses $1+2$ and $3+4$ interfering. One of the output ports is monitored by the classical detector, which now measures both $\mu_H$ and $\mu_V$ depending on the time stamp. These output pulses are brought again to another interferometer where one arm is in H polarization and the other is in V polarization. Here the interference of $1+2+3+4$ pulses is what we are looking for. Of course, interference from other time bins are present too (such as $2+3+4+5$, $3+4+5+6$, etc.) and to prevent leakage we should use a fixed intensity modulator to suppress every three out of four output pulses. Note that the intensity modulator only works like an optical switch that selects one out of every four signals following a given pattern, so it does not contain random numbers and does not leak additional information.}
	\label{fig:setup_simple}
\end{figure}

An alternative setup, inspired by a similar delay-line-based setup in Ref. \cite{passivedecoy2}, is shown in Fig. \ref{fig:setup_simple}. Here instead of using four independent light sources, Alice simply uses one single source, and employs delay lines to interfere four signals from different time bins, which ensures that the pulses will interfere with good visibility. Also, importantly, no phase stabilization is required between the upper/lower arms for the two Mach-Zehnder interferometers (since we only need the relative phases between pulses to be random, and also since we are measuring at the output ports, any internal phase drift simply adds on to the random number).

There are two points worth noting if one uses such a setup: (1) as mentioned in assumption 1, one need to ensure that neighboring pulses from the same source do not have phase correlations; (2) here only one event out of four time bins is used for the actual source state generation, but the other discarded events contain partial information of the phase, so one needs to suppress these output signals with an active intensity modulator (IM). This does not break the premise of the passive source, as this intensity modulator follows a fixed pattern (letting pass one in every four pulses) and does not contain any encoding information. Note that, though, in practice the intensity modulator would only have finite extinction ratio, which may lead to leakage of some partial information about the phases to Eve. This problem would be a subject of future studies. 

Another alternative solution is to use an optical switch to replace the intensity modulator, although it might be slower than IMs. An additional note here is that the setup of Fig. \ref{fig:setup_simple} requires polarization independence, which may be a challenge for the IM. Some optical switches like acoustic-optical modulators (AOMs) have negligible polarization dependence, although they might be slower than the electro-optical counterparts.

\section{Simulation Results}

\begin{table}
	\begin{tabular}{cccc}
	    $p_d$  & $\eta_d$  & $f_e$ & $\epsilon$\\
		\hline
		$10^{-6}$& $1$ & $1.16$ & $10^{-7}$\\
	\end{tabular}
	\caption{List of fixed parameters used for the simulation. $p_d$ is the per-detector dark count. The detector efficiency $\eta_d$ is merged into channel loss and is therefore set to 1. $f_e$ is the error-correction efficiency and $\epsilon$ is the failure probability for the finite-size effect (corresponding to a confidence interval $\gamma=5.3$ number of standard deviations from the mean value).}
\end{table}

\begin{figure}[h]
	\includegraphics[scale=0.175]{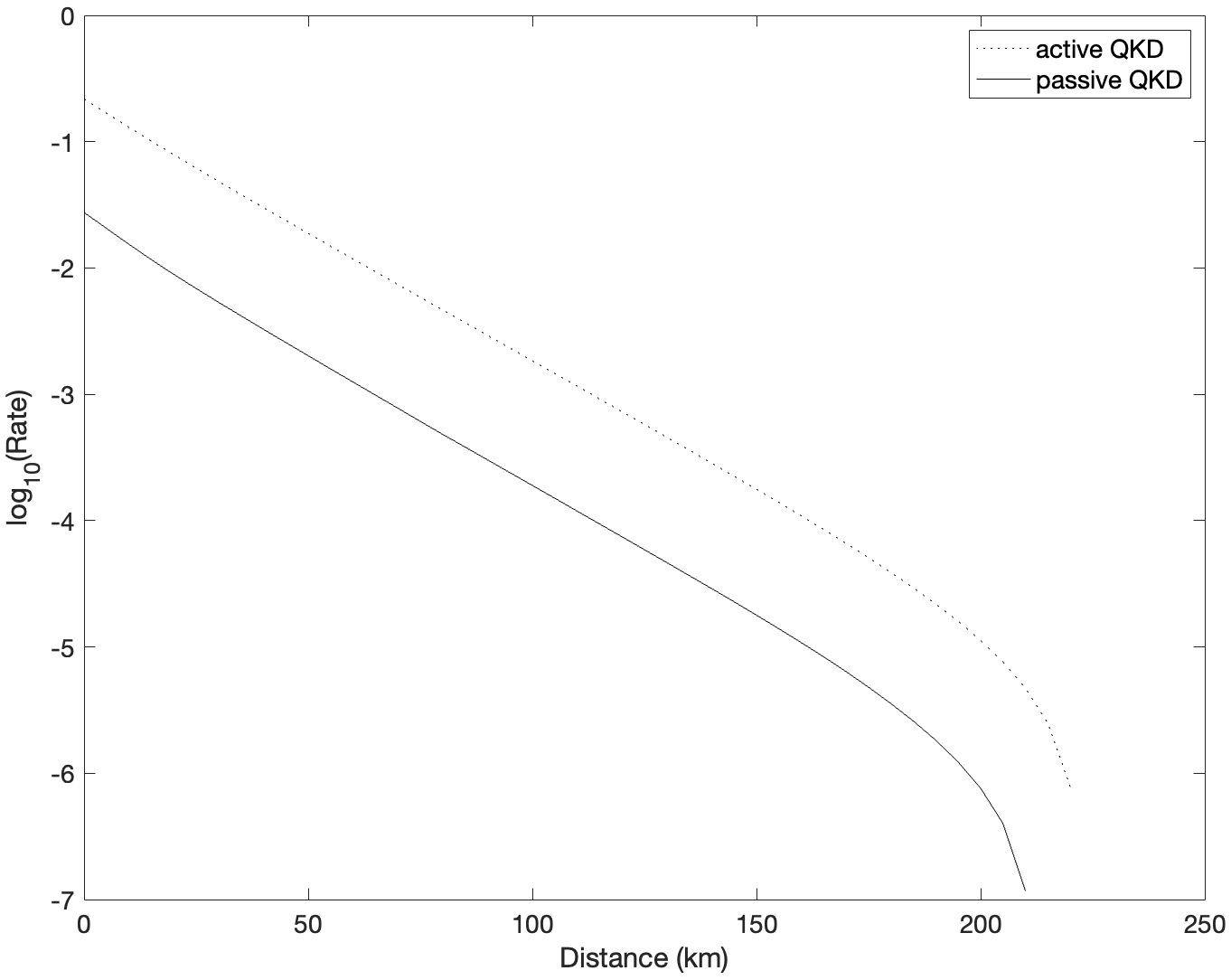}
	\caption{Key rate comparison between active BB84 and fully passive BB84. Parameters from Table I are used, and the misalignment is set to $e_d=2\%$ on the X-Z plane. Here an infinite data size is assumed and 3 decoy regions are used. We optimize the Z (key generation) basis post-selection threshold $\Delta_Z$ (which takes a value between $[0.05,0.1]$) and the maximum intensity $\mu_{max}$ which takes a value $[0,1]$. The X/Y bases slice sizes and the decoy state thresholds are chosen to be $\Delta_{XY}=0.1,\Delta_\phi=0.1,t_{decoy}=0.04/\mu_{max},t_{decoy2}=0.02/{\mu_{max}}$ (as the asymptotic scenario always favors smaller decoy intensities and smaller slices, we just set them to be fixed to some small and reasonable values). We can see that the passive setup has a reasonable key rate, although it is about one order-of-magnitude lower than the active counterpart. The difference mainly comes from the trade-off between signal QBER and sifting probability.}
	\label{fig:BB84rate}
\end{figure}

\begin{figure}[h]
	\includegraphics[scale=0.175]{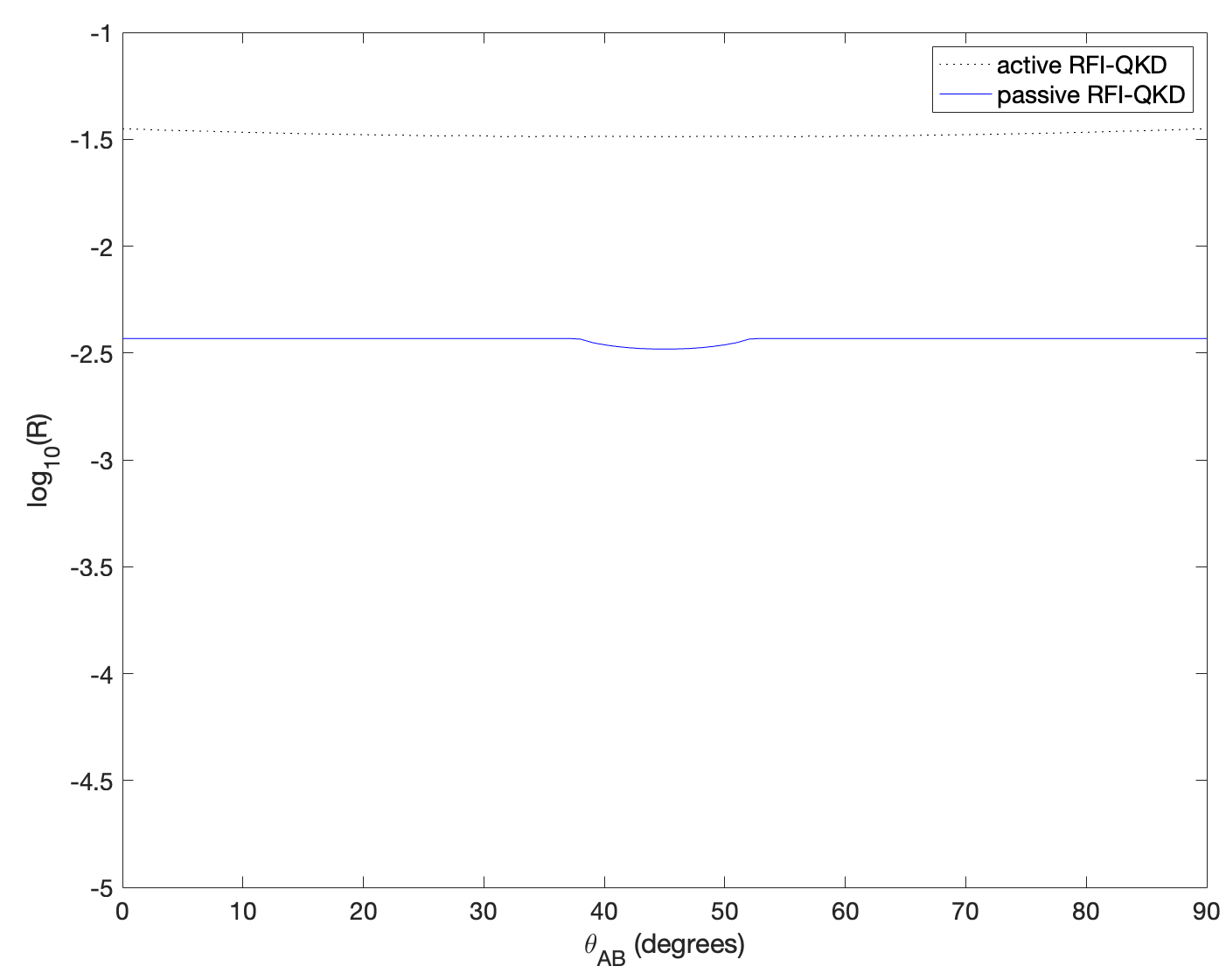}
	\caption{Key rate comparison between active RFI-QKD and fully passive RFI-QKD at $L=50km$, plotted against a rotation along the $Z$ axis (i.e. along the X-Y plane) of angle $\theta_{AB}$. Parameters from Table I are used, and $\Delta_Z, \Delta_{XY}$ and $\Delta_{\phi}$ are all fixed to $0.1$ rad. Here an infinite data size is assumed and 3 decoy regions are used, with $\mu_{max}=1,t_{decoy}=0.04,t_{decoy2}=0.02$. As can be seen, passive RFI-QKD can provide the same resilience against rotation in the X-Y plane like its active counterpart, but again with about one order-of-magnitude lower key rate due to post-selection in the Z basis. A slight dip (imperfection) in the middle is likely caused by the finite size of the post-selection slice, as the dip disappears when $\Delta_{\phi}$ is set to a smaller value. This is likely because a finite phase slice is equivalent to a rotation angle changing with time in RFI-QKD, which the analysis cannot fully correct.}
	\label{fig:RFIQKD}
\end{figure} 

\begin{figure}[h]
	\includegraphics[scale=0.175]{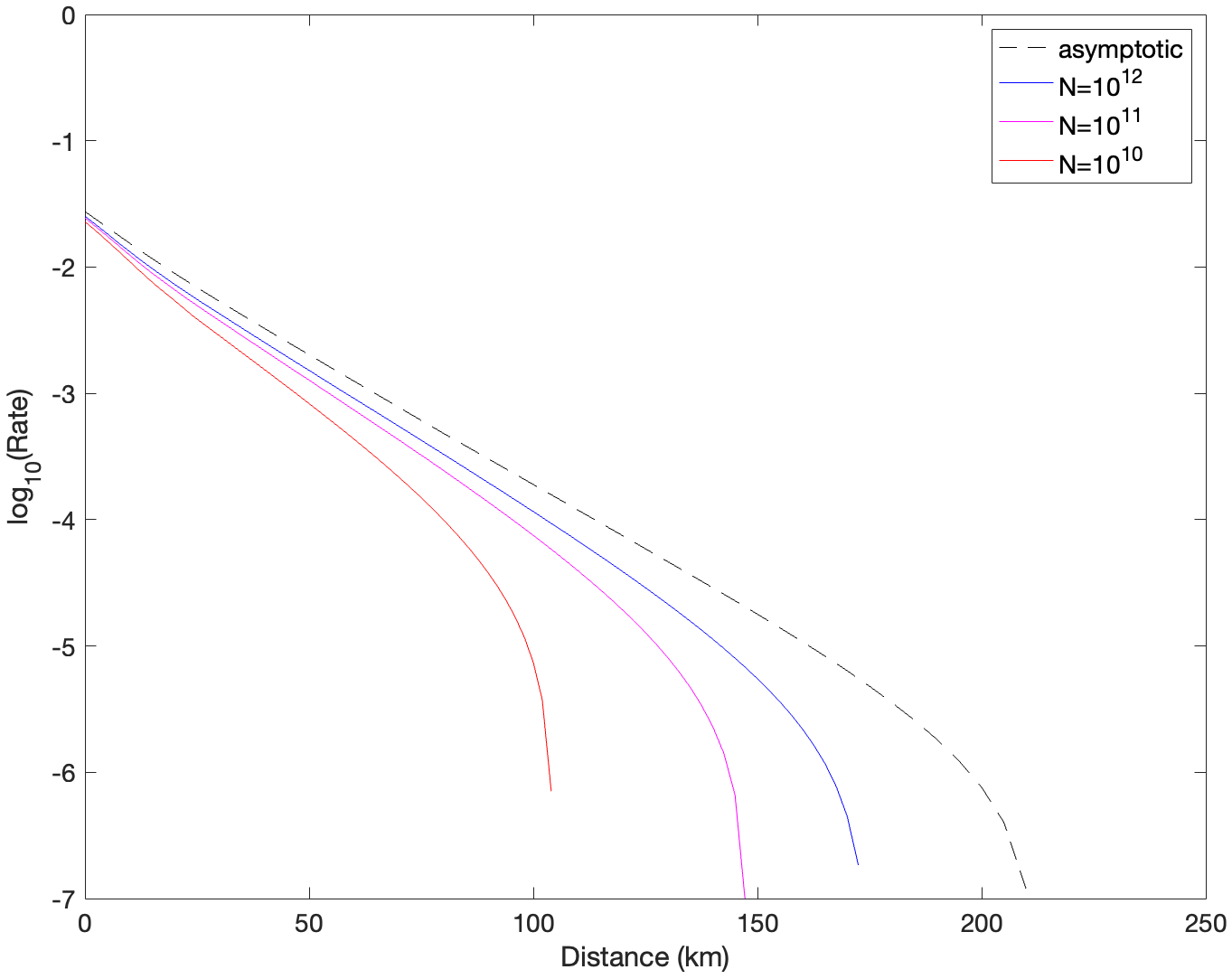}
	\caption{Key rate simulation for passive BB84 under finite-size effects (considering collective attacks) with different data sizes. For reference we also plot the asymptotic case with an infinite data size and 3 decoy regions. Here the misalignment is set to $e_d=2\%$ on the X-Z plane. We use the system parameters from Table I, and perform full optimization on all post-selection parameters, including $\mu_{max},\Delta_Z,\Delta_{XY},\Delta_\phi,t_{decoy}$ and $t_{decoy2}$, as well as Bob's probability for choosing the Z basis. We can see that even with relatively small data size like $N=10^{10}$ of transmitted signals, we can still get an acceptable key rate, despite the penalty from statistical fluctuation and from post-selection in the passive source. }
	\label{fig:BB84ratefinite}
\end{figure} 

In this section we simulate the key rate for QKD with our fully passive source, specifically for two protocols, BB84 and RFI-QKD.

In Fig. \ref{fig:BB84rate}, we plot the key rate for BB84 (with 3 decoy settings and infinite data size) where active versus passive sources are used. We can see that a reasonable key rate can be obtained despite the post-selection process applied to the source and the inherent QBER from the finite slices. Due to the sifting, there is still a decrease of around one order-of-magnitude consistently across different distances, though, in exchange for the more secure implementation of the QKD source.

In Fig. \ref{fig:RFIQKD}, we similarly plot a comparison between the achievable key rate of using active versus passive sources for RFI-QKD. Here we fix the distance to $50km$ and simulate the key rate at different rotation angles between 0 to 90 degrees on the Bloch sphere (angles above 90 degrees will yield symmetric results so they are not shown here). We can see that we can still implement RFI-QKD and acquire resilience against rotation along the Z axis even when using the passive source. There is no additional sifting penalty to the key rate compared to passive BB84 since we are preparing the Y basis states to begin with (and simply use these data for RFI-QKD, which would have been discarded for passive BB84). Note that there is a small decrease in the key rate depending on the misalignment angle, likely due to the finite slices $\Delta_{XY}$ and $\Delta_\phi$. For instance, $\Delta_\phi$ has a similar effect to a changing misalignment angle (or a depolarization), which is no longer a unitary rotation and cannot be corrected completely by the RFI analysis, which requires the angle to have an unknown but fixed value.

Lastly, we also tested passive BB84 under finite-size effects considering collective attacks. From Fig. \ref{fig:BB84ratefinite} we can see that we can still maintain an acceptable key rate for fully passive BB84 at reasonable data sizes such as $10^{11}$ or $10^{12}$. 

For conciseness, we have only considered BB84 and have not considered RFI-QKD with respect to the finite-size here, but in principle, if we only consider collective attacks, a similar approach can be taken to estimate the fluctuations in the $XX,XY,YX,YY$ and $ZZ$ gains and error-gains and one can simply take the worst-case values for each of these single-photon QBER values.

\section{Discussion}
	
In this work we have presented a simple yet effective scheme to implement a fully passive QKD source that performs both decoy state choice and encoding by post-selection, which makes the passive scheme practical for the first time. To demonstrate its practical performance, we have also explained in theory how to use it to implement the BB84 and RFI-QKD protocols and have shown that we can obtain reasonable key rates, albeit slightly lower than the active case due to sifting, which is the price for obtaining better security and avoiding side-channels in modulators. 

Our source is in principle also compatible with MDI-QKD, and a similar post-selection idea on phases can even be applied to Twin-Field QKD sources. Such protocols, combined with passive sources, will have highly desirable implementation security as they eliminate side-channels from both the source modulators and the detectors. A detailed study of their setup and performance analysis would be the subject of our future studies. 

\section{Acknowledgments}

We thank Yue Jiang, Kai Sum Chan, Chengqiu Hu, and Chenyang Li for helpful discussions. The project is supported by NSERC, MITACS, CFI, ORF, Connaught Innovation Award, US Office of Naval Research, Huawei Technologies Canada, Inc., the Royal Bank of Canada, and the University of Hong Kong start-up grant. 

M.C. acknowledges support from the Galician Regional Government (consolidation of Research Units: AtlantTIC), the Spanish Ministry of Economy and Competitiveness (MINECO), the Fondo Europeo de Desarrollo Regional (FEDER) through Grant No. PID2020-118178RB-C21, and MICIN with funding from the European Union NextGenerationEU (PRTR-C17.I1) and the Galician Regional Government with own funding through the “Planes Complementarios de I+D+I con las Comunidades Autonomas” in Quantum Communication.

\appendix

\section{Mapping between Source and Output States} \label{ref:appendix_mapping}

In this Appendix we describe how can one convert the four degrees-of-freedom in the input signal phases $\{\phi_1,\phi_2,\phi_3,\phi_4\}$ to the states characterized by $\{\mu_H,\mu_V,\phi_H,\phi_V\}$ and eventually to the output state described by four degrees-of-freedom $\{\mu,\theta_{HV},\phi_{HV},\phi_G\}$.

As we described in the main text, two interfering coherent lights characterized by $(\mu_1,\phi_1)$ and $(\mu_2,\phi_2)$ will end up with the output state:

\begin{equation}
	\begin{aligned}
		&\ket{\sqrt{\mu_1} e^{i\phi_1}}_a \ket{\sqrt{\mu_2} e^{i\phi_2}}_b \rightarrow \\
		&\ket{\sqrt{\mu_1/2} e^{i\phi_1}+i\sqrt{\mu_2/2} e^{i\phi_2}}_c \ket{i\sqrt{\mu_1/2} e^{i\phi_1}+\sqrt{\mu_2/2} e^{i\phi_2}}_d.
	\end{aligned}
\end{equation}

If we only look at port c, we can consider the state $\ket{\sqrt{\mu_1/2} e^{i\phi_1}+i\sqrt{\mu_2/2} e^{i\phi_2}}_c$ as a vector on the complex plane, as shown in Fig. \ref{fig:state}, on which two input vectors of lengths $\sqrt{\mu_1/2},\sqrt{\mu_2/2}$ at polar angles $\phi_1,\phi_2+\pi/2$ are added up (the second angle is shifted by $\pi/2$ to account for the factor $i$). The new state (which has polarization H) can be thought of as a coherent state with amplitude $\mu_H$ and global phase $\phi_{H}$ as:

\begin{equation}
	\begin{aligned}
		\mu_H &= \mu_1/2 + \mu_2/2 + \cos(\phi_2 + \pi/2 - \phi_1) \sqrt{\mu_1\mu_2}, \\
		\phi_H &= \phi_1 + \sin^{-1}(\sqrt{\mu_1/(2\mu_H)}\sin(\phi_2 + \pi/2 - \phi_1)).
	\end{aligned}
\end{equation}

For simplicity, here we denote $\phi_2' = \phi_2 + \pi/2$, and consider $\mu_1=\mu_2$, then

\begin{equation}
	\begin{aligned}
		\mu_H &= \mu_1 [1 + \cos(\phi_2' - \phi_1)] ,\\
		\phi_H &= (\phi_1 + \phi_2')/2.
	\end{aligned}
\end{equation}

Similarly, if $\mu_3 = \mu_4$, we have that

\begin{equation}
	\begin{aligned}
		\mu_V &= \mu_3 [1 + \cos(\phi_4' - \phi_3)] ,\\
		\phi_V &= (\phi_3 + \phi_4')/2.
	\end{aligned}
\end{equation}

Here we have obtained the two intermediate coherent states described by $\{\mu_H,\mu_V,\phi_H,\phi_V\}$. Next the two states are joined at the PBS. Using the fact that coherent states satisfy:

\begin{equation}
	\begin{aligned}
		\ket{\alpha} = e^{-|\alpha|^2/2} e^{\alpha a^\dagger} e^{-\alpha^* a} \ket{vac} \\
	\end{aligned}
\end{equation}

\noindent we can write the output state as:

\begin{equation}
	\begin{aligned}
		&\ket{\sqrt{\mu_H} e^{i\phi_H}}_H \ket{\sqrt{\mu_V} e^{i\phi_V}}_V\\
		=& e^{-\mu_H/2}e^{\sqrt{\mu_H}e^{i\phi_H} a_H^\dagger}e^{-\sqrt{\mu_H}e^{-i\phi_H} a_H} \\
		&\times e^{-\mu_V/2}e^{\sqrt{\mu_V}e^{i\phi_V} a_V^\dagger}e^{-\sqrt{\mu_V}e^{-i\phi_V}a_V} \ket{vac}\\
		=& e^{-(\mu_H+\mu_V)/2} e^ {\left( {\sqrt{\mu_H}e^{i\phi_{H}}a_H^\dagger + e^{i\phi_{V}}\sqrt{\mu_V}a_V^\dagger}\right)}\\
		&\times e^{\left({-\sqrt{\mu_H}e^{-i\phi_{H}}a_H - e^{-i\phi_{V}}\sqrt{\mu_V}a_V} \right)}\ket{vac} \\
		=& e^{-(\mu_H+\mu_V)/2} \\ &\times e^{\sqrt{\mu_H+\mu_V}e^{i\phi_{H}}\left({\sqrt{\mu_H\over{\mu_H+\mu_V}}}a_H^\dagger + {\sqrt{\mu_V\over{\mu_H+\mu_V}}}e^{i(\phi_V-\phi_H)}a_V^\dagger \right)} \\
		&\times e^{-\sqrt{\mu_H+\mu_V}e^{-i\phi_{H}}\left({\sqrt{\mu_H\over{\mu_H+\mu_V}}}a_H + {\sqrt{\mu_V\over{\mu_H+\mu_V}}}e^{-i(\phi_V-\phi_H)}a_V \right)} \\
		&\ket{vac} \\
		=& \ket{ \sqrt{\mu_H+\mu_V}e^{i\phi_H} }_{\theta_{HV},\phi_{HV}},
	\end{aligned}
\end{equation}

\noindent which is still a coherent state, but in a new polarization state defined by $\theta_{HV}$ and $\phi_{HV}$, with a global phase $\phi_H$. We define the creation operator in the new polarization mode in Eq. \ref{eq:polarization}, which describes the polarization in terms of $\theta_{HV}$ and $\phi_{HV}$, and the photon number states in the given polarization mode can be written as Eq. \ref{eq:photon_number}.

This means that we can describe the output state with four degrees-of-freedom: the intensity $\mu=\mu_H+\mu_V$, the polarization on the Bloch sphere (which contains two degrees-of-freedom $\theta_{HV}$ and $\phi_{HV}$), and the overall global random phase $\phi_G = \phi_H$. That is, any combination of $\{\phi_1,\phi_2,\phi_3,\phi_4\}$ can be mapped to one pair of coherent states described by $\{\mu_H,\mu_V,\phi_H,\phi_V\}$ and then to a single output coherent state described by parameters $\{\mu,\theta_{HV},\phi_{HV},\phi_G\}$.

Inversely, for any given state described by $(\theta_{HV},\phi_{HV})$ on the Bloch sphere and with intensity $\mu \in [0,\mu_{max}]$ (here we assume that all the four input signals have identical intensities $\mu_1=\mu_2=\mu_3=\mu_4$, in which case $\mu_{max}=\mu_1+\mu_2=\mu_3+\mu_4$) and phase $\phi_G$, we can uniquely obtain

\begin{equation}
	\begin{aligned}
		\phi_H &= \phi_G ,\\
		\phi_V &= \phi_{HV} + \phi_G ,\\
		\mu_H &= \mu \cos^2(\theta_{HV}/2),\\
		\mu_V &= \mu \sin^2(\theta_{HV}/2).\\
	\end{aligned}
\end{equation}

For any given set $\{\mu_H,\phi_H\}$ where $\mu_H \in [0,\mu_1+\mu_2]$, assuming $\mu_1=\mu_2$, we can solve for two symmetric solutions (or one solution if $\mu_H=0$ or $\mu_H=\mu_1+\mu_2$) for $\phi_1$ and $\phi_2$:

\begin{equation}
	\begin{aligned}
		\phi_1 &= \phi_H \pm \cos^{-1} (\mu_H/\mu_1 - 1)/2 ,\\
		\phi_2 &= \phi_H \mp \cos^{-1} (\mu_H/\mu_1 - 1)/2\\
	\end{aligned}
\end{equation}

\noindent and a similar argument applies to the V state.

Up to here we have shown that, any set of input states with phases $\{\phi_1,\phi_2,\phi_3,\phi_4\}$ can always be mapped to an output coherent state described by the parameters $\{\mu,\theta_{HV},\phi_{HV},\phi_G\}$. Inversely, for any output state characterized by $\{\mu,\theta_{HV},\phi_{HV},\phi_G\}$, given that $\mu\leq \mu_{max}$, we can at least find one set of input states that can generate this output. This means that, with the fully passive source, we can in principle create a coherent state with arbitrary intensity, and with its creation operator describing any polarization state (or a state in other encoding degrees-of-freedom such as time-bin phase encoding or path encoding). Combined with the decoy-state analysis, this allows us to simulate a source that can output any arbitrary state of a qubit system.

\section{Channel Model} \label{ref:appendix_channel}

In this Appendix we briefly describe the channel models used in the simulations to calculate the observed quantities $\langle Q \rangle_{S_i}$ and $\langle QE \rangle_{S_i}$ based on a post-selection region $S_i$.

In Sec. III. D when describing decoy states, we have focused on the intensities $\mu_H$ and $\mu_V$, but in fact the most general description of an observable is a function of $(\mu_H,\mu_V,\phi)$, which uniquely defines a point on a Bloch sphere. The expected value of this function can be written as

\begin{equation}
	\begin{aligned}
		&\langle f(\mu_H,\mu_V,\phi_{HV}) \rangle_{S_i} \\
		&= \iiint_{S_i} p(\mu_H,\mu_V,\phi_{HV})f(\mu_H,\mu_V,\phi_{HV}) d\mu_H d\mu_V d\phi_{HV} \\
		&\times \left( 1 / \iiint_{S_i} p(\mu_H,\mu_V,\phi_{HV})d\mu_H d\mu_V d\phi_{HV} \right),
	\end{aligned}
\end{equation}

\noindent where the actual post-selection regions $\{S_i\}$ are defined in Section III.\\

Now, to establish a channel model and simulate the observed statistics in an experiment, our goal is to write out the gain and QBER for any given set of $(\mu_H,\mu_V,\phi_{HV})$. That is, we need to derive the functions $Q(\mu_H,\mu_V,\phi_{HV})$ and $QE(\mu_H,\mu_V,\phi_{HV})$ for each pair of bases. We can do this with two steps: characterizing the source and characterizing the channel.\\

1. Firstly, let us characterize the state being sent. The output state of the PBS (which combines input signals of intensities $\mu_H$ and $\mu_V$ and relative phase $\phi_{HV}$) is a coherent state

\begin{equation}
	\ket{\sqrt{\mu}}_{\theta_{HV},\phi_{HV}} = \sum_{n=0}^{\infty} \sqrt{P_n} \ket{n}_{\theta_{HV},\phi_{HV}}
\end{equation}

\noindent where $P_n$ is the Poissonian distribution defined by $\mu=\mu_H+\mu_V$, and the polarization is determined by the ratio of $\mu_H/\mu_V$ as well as the relative phase $\phi_{HV}$. More specifically, the creation operator of the polarization mode is described by parameters $\theta_{HV}$ and $\phi_{HV}$, as shown in Eq. \ref{eq:creation_operator}.

The polarization of each single photon can be described as a vector on the Bloch sphere with coordinates:

\begin{equation}
	\begin{aligned}
		\vec{s} &= (\sin\theta_{HV} \cos\phi_{HV}, \sin\theta_{HV} \sin\phi_{HV}, \cos\theta_{HV}).
	\end{aligned}
\end{equation}\\

2. Then, let us characterize the channel (and Bob's detection system) by four elements: misalignment $e_d$, loss $\eta$, detection efficiency $\eta_d$, and dark count rate $p_d$ of the detectors. The element of particular interest here is the misalignment. Previous works usually characterize $e_d = \sin^2 (\alpha)$ as a simple rotation along the X-Z plane that unitarily rotates the states as:
\begin{equation}
	\begin{aligned}
		\ket{H} &\rightarrow \cos\alpha\ket{H} + \sin\alpha\ket{V} ,\\
		\ket{V} &\rightarrow -\sin\alpha\ket{H} + \cos\alpha\ket{V} .\\
	\end{aligned}
\end{equation}

However, in a fibre channel, in fact the rotation could be along any axis on the Bloch sphere, which could be defined by a unit axis vector $\vec{r}=(r_X,r_Y,r_Z)$ and a rotation angle $\alpha$. The above X-Z rotation can simply be defined as an angle of $\alpha$ around the axis $(0,1,0)$.

Using the Rodrigues' rotation formula \cite{Rodrigues}, each single photon in the state $\vec{s}$ will be rotated into a state

\begin{equation}
	\vec{s'} = \cos\alpha \vec{r} + \sin\alpha (\vec{s} \times \vec{r}) + (\vec{v} \cdot \vec{r})(1-\cos\alpha)\vec{s}.
\end{equation}

Afterwards, the projection probability of the new state onto a detection basis $\vec{b}=(b_X,b_Y,b_Z)$ will be

\begin{equation}
	\begin{aligned}
		P_{\text{proj}} &= \cos^2(\alpha_{\text{proj}}/2).
	\end{aligned}
\end{equation}

\noindent where $\alpha_{\text{proj}}$ is defined as

\begin{equation}
	\begin{aligned}
		\alpha_{\text{proj}} &= \cos^{-1}(\vec{s'} \cdot \vec{b}) ,\\
	\end{aligned}
\end{equation}

{\color{black} We can then calculate the probabilities of clicking for each pair of detectors if Alice sent a coherent state with intensity $\mu$.} Here $0$ corresponds to the detected state along the basis $\vec{b}$, and $1$ corresponds to the opposite state $-\vec{b}$. For instance, H corresponds to projecting onto $(0,0,1)$ and V to $(0,0,-1)$:

\begin{equation}
	\begin{aligned}
		P_{\text{click}}^0 &= 1 - (1-p_d)e^{-\eta\eta_d\mu P_{\text{proj}}} ,\\
		P_{\text{click}}^1 &= 1 - (1-p_d)e^{-\eta\eta_d\mu (1-P_{\text{proj}})}.
	\end{aligned}
\end{equation}

\noindent where we have also included channel loss $\eta$ and detector efficiency $\eta_d$.

Based on the state prepared by Alice and the detector event obtained by Bob, we can easily calculate the average gain and error-gain $\langle Q_j \rangle_{S_i}$ and $\langle QE_j \rangle_{S_i}$ for the basis $i$ prepared by Alice and the basis $j$ measured by Bob.

\section{Addressing Incompatibility of Decoy State Analysis with Passive Source}\label{ref:appendix_decoy}

\begin{figure*}[t]
	\includegraphics[scale=0.2]{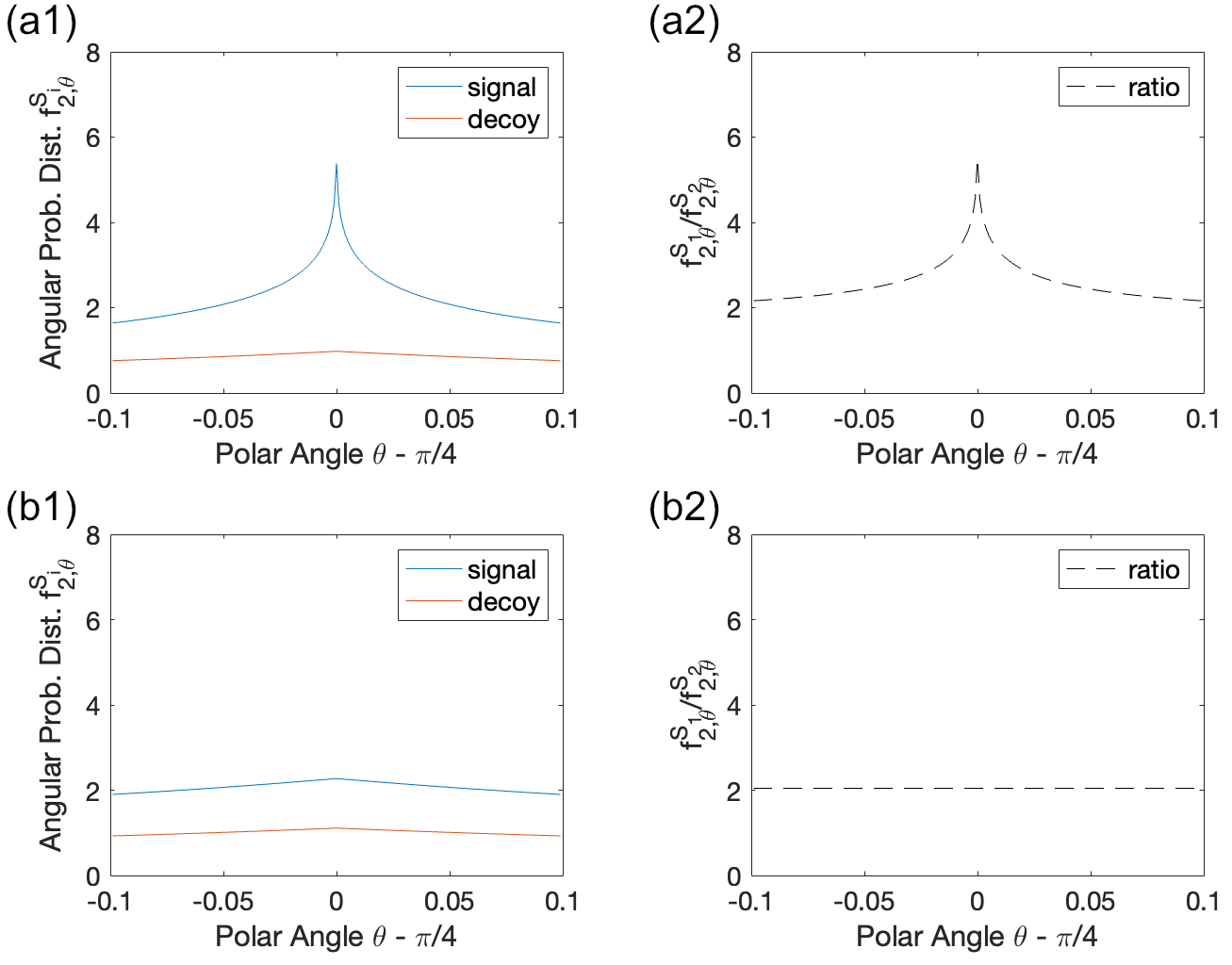}
	\caption{Example of polar probability distributions $f_{n,\theta}^{S_i}$ for the X and Y basis states, which are the photon number distribution multiplied by the intensity distribution, partially integrated over the radial direction and leaving only an angular dependency. $\theta$ is the deviation of the polarization from the diagonal $\theta=\pi/4$ line (i.e. perfect X or Y basis state). Here we show the 2-photon polar probability distributions, i.e. $f_{2,\theta}^{S_i}$, for two different decoy settings $S_1$ and $S_2$ (signal and decoy states). (a1): Polar probability distributions using the ``naive" strategy. (a2) The ratio of the two distributions for two decoy settings. As we can observe, the distributions are not proportional for the two decoy settings, hence no linear program can be constructed for arbitrary $Y_2(\theta)$. (b1) Polar probability distributions using the sector-shape post-selection regions and post-selected intensity distribution $exp(\mu_H+\mu_V)$. (b2) The ratio of the two distributions for two decoy settings. As can be seen, now the probability distributions for the two decoy settings have exactly the same shape and are proportional to each other, allowing us to perform the decoy analysis on the same set of variables $(\int f_{n,\theta}^{S_i}(\theta) Y(\theta) d\theta)/(\int f_{n,\theta}^{S_i}(\theta) d\theta)$. Note that here we have normalized all $f_{n,\theta}$ with $1/P_{S_i}$ in order to improve the plot visibility. This does not affect the conclusion that for the ``naive" strategy the polar probability distributions are not proportional and for our proposed strategy they are.}
	\label{fig:photondist_new}
\end{figure*}

In this Appendix we discuss in more detail of the incompatibility of the decoy-state analysis with passive source, as well as our solution to address this issue.

Here, we propose a \textit{necessary condition} for the decoy-state analysis to be valid, as a simple test criteria for whether a given strategy is compatible with it.

For the decoy-state analysis (i.e. the linear program) to hold true, we must have a consistent set of variables that are independent of the decoy choice $S_i$, valid for all $n$. First let us consider active QKD. Here the set of consistent variables are the yields ``$Y_n$" (or $e_nY_n$, but here we focus on the yields). In other words, for active QKD, two terms corresponding to the n-photon contributions satisfy

\begin{equation}
	\begin{aligned}
		{{P_n^{\mu}Y_n}\over{P_n^{\nu}Y_n}} = {{P_n^{\mu}}\over{P_n^{\nu}}} 
	\end{aligned}
\end{equation}

\noindent which is a constant ${{P_n^{\mu}}/{P_n^{\nu}}} = ({{\mu^n}/{\nu^n}}) exp(-\mu+\nu)$. Importantly, it is independent of $Y_n$. These constant known coefficients allow us to construct a linear program with a set of $Y_n$ as the unknown variables across various linear constraints. That is each pair of n-photon contribution terms associated with different decoy settings having a constant ratio, independent of $Y_n$, is a necessary condition for the decoy-state analysis to be valid.\\

Now, let us consider the passive QKD case. As we mentioned in the main text, the terms in the linear constraints $Q_{S_i}=\sum_n \langle P_n Y_n \rangle_{S_i} $ are in the form of:

\begin{equation}
	\begin{aligned}
		\langle P_n Y_n \rangle_{S_i} &= {1\over P_{S_i}}\iint_{S_i} P_n(r,\theta) p(r,\theta) Y_n(\theta) rdrd\theta \\
		 &= {1\over P_{S_i}}\iint_{S_i} f_n(r,\theta) Y_n(\theta) drd\theta, \\
	\end{aligned}
\end{equation}

\noindent where we denote $f_n(r,\theta)=P_n(r,\theta) p(r,\theta)r$ as the overall probability distribution for convenience.

A necessary condition here for the successful construction of a linear program is therefore that the ratio of the n-photon contributions for two decoy settings $S_i$ and $S_j$
\begin{equation}
	\begin{aligned}
		{{\langle P_n Y_n \rangle_{S_i} } \over {\langle P_n Y_n \rangle_{S_j}}} 
		&= {P_{S_j}\over P_{S_i}} { {\iint_{S_i} f_n(r,\theta) Y_n(\theta) drd\theta} \over  {\iint_{S_j} f_n(r,\theta) Y_n(\theta) drd\theta} }, \\
	\end{aligned}
\end{equation}

\noindent must be a constant value independent of the unknown function $Y_n(\theta)$.\\

To reduce the problem to a one-dimensional problem (which helps us visualize it), we can partially integrate the above overall probability distribution over $r$, to obtain

\begin{equation}
	\begin{aligned}
		\langle P_n Y_n \rangle_{S_i} &= {1\over P_{S_i}}\int_{\theta_{min}}^{\theta_{max}} f_{n,\theta}^{S_i}(\theta) Y_n(\theta) d\theta, \\
	\end{aligned}
\end{equation}

\noindent where

\begin{equation}
	\begin{aligned}
		f_{n,\theta}^{S_i}(\theta) &= \int_{r_{min,i}(\theta)}^{r_{max,i}(\theta)} f_n(r,\theta)dr. \\
	\end{aligned}
\end{equation}

Here we denote the distribution $f_{n,\theta}^{S_i}(\theta)$ as the ``polar probability distribution", which is the overall probability distribution (including the Poissonian distribution for the photon number $P_n$ and the intensity distribution $p_\mu$) partially integrated over the radial direction first, i.e. the remaining distribution is the angular probability distribution. 

The ratio of the n-photon contributions to the observables for two decoy settings is now

\begin{equation}
	\begin{aligned}
		{{\langle P_n Y_n \rangle_{S_i}}\over{\langle P_n Y_n \rangle_{S_j}}} = {{P_{S_j}}\over{P_{S_i}}} {{\int_{\theta_{min,i}}^{\theta_{max,i}} f_{n,\theta}^{S_i}(\theta) Y_n(\theta) d\theta }\over{\int_{\theta_{min,j}}^{\theta_{max,j}} f_{n,\theta}^{S_j}(\theta) Y_n(\theta) d\theta }}.
	\end{aligned}
\end{equation}

Importantly, just like how $Y_n$ is a fixed variable independent of the decoy setting in active QKD, here for our passive source scenario, $Y_n(\theta)$ is independent of the decoy setting. It is a fixed yet \textit{unknown} function (subject to Eve's control) that is the same for all $S_i$. To make the above expression \textit{always} a constant that is independent of $Y_n(\theta)$ \footnote{We choose the ideal case where $S_i$ and $S_j$ have the same angular integral region $\theta_{min},\theta_{max}$. If $S_i$ and $S_j$ have different ranges of $\theta$, it is likely that the ratio cannot be independent of $Y_n(\theta)$.}, the only viable solution is to impose that the angular probability distributions satisfy

\begin{equation}\label{eq:ratio}
	\begin{aligned}
		f_{n,\theta}^{S_i}(\theta)/f_{n,\theta}^{S_j}(\theta) = C_{S_i,S_j,n},
	\end{aligned}
\end{equation}

\noindent where the ratio is a constant $C_{S_i,S_j,n}$. That is, the two angular probability distribution functions are proportional to each other. \footnote{A simple intuitive proof for the above Eq. \ref{eq:ratio} is the following: suppose two integrals $\int_{\theta_{min}}^{\theta_{max}} f_1(\theta) Y(\theta) d\theta$  and $\int_{\theta_{min}}^{\theta_{max}} f_2(\theta) Y(\theta) d\theta$ are always equal for every possible distribution $Y(\theta)$, then one can consider the case when $Y=\delta(\theta-\theta_0)$, which is a delta function centered at $\theta_0$, a constant value. This means that we must have $f_1(\theta_0)=f_2(\theta_0)$. However, since $Y(\theta)$ can be any function, $\theta_0$ can be any value between $\theta_{min} \leq \theta_0 \leq \theta_{max}$. We therefore find that $f_1(\theta)=f_2(\theta)$ across the entire region $[\theta_{min},\theta_{max}]$. A similar argument applies if the two integrals differ by a constant ratio, in which case $f_1$ and $f_2$ also differ by a constant ratio.}

Now, the problem is, for any arbitrary choice of integral region $S_i$ and the intensity probability distribution $p_\mu$, generally speaking, $f_{n,\theta}(\theta)$ will be neither identical nor proportional functions for different sets of the integration region $S_i$, i.e. the decoy setting. This means that, a naive implementation of the passive source and the post-selection regions will not be compatible with the decoy-state analysis. We illustrate an example of this in Fig. \ref{fig:photondist_new} (a1), (a2), where we can see that the angular probability distributions are not proportional, meaning that a linear program cannot be constructed.\\

For our solution, we apply an additional post-selection $q_\mu$ on the intensity distribution, such that the overall probability distribution satisfies

\begin{equation}
	\begin{aligned}
			f_n(r,\theta) &= P_n(r,\theta)q_\mu(r,\theta)p_\mu(r,\theta)r \\
			&= r^{n+1}(\sin\theta+\cos\theta)^n/n!,\\
	\end{aligned}
\end{equation}

\noindent which can be decoupled into the direct product of a radial part and an angular part.

Moreover, we select integral regions $S_i$ that are concentric sector-shapes, which can be converted into rectangular regions on the $(r,\theta)$ domain: $(\theta_{min},\theta_{max})$ and $(0,r_{max,i})$, where the only difference for the regions are the radius $r_{max,i}$.

Now, the angular probability distributions are simply

\begin{equation}
	\begin{aligned}
		f_{n,\theta}^{S_i}(\theta) &= {{r_{max,i}^{n+2}}\over{(n+2) n!}} \times (\sin\theta + \cos\theta)^n ,\\
	\end{aligned}
\end{equation}

\noindent such that 

\begin{equation}
	\begin{aligned}
		f_{n,\theta}^{S_1}(\theta)/f_{n,\theta}^{S_2}(\theta) = (r_{max,1}/r_{max,2})^{n+2},
	\end{aligned}
\end{equation}

\noindent which is a known constant. This means that, with the new post-selected intensity distribution and the specifically chosen $S_i$, it is now possible to construct a consistent set of variables for the linear programs, as the terms ${\langle P_n Y_n \rangle_{S_1}}$ and ${\langle P_n Y_n \rangle_{S_2}}$ only differ by a known constant ratio, which constitutes the coefficients for the variables in the linear program, just like what we have for active QKD. We illustrate an example of this in Fig. \ref{fig:photondist_new} (b1) and (b2), where we can see that the angular probability distributions are proportional, hence the integral with any $Y_n(\theta)$ is also proportional, meaning that a linear program can be constructed.

Note that, importantly, the above strategy is not the only solution. For instance, as long as we still choose the same sector-shape integral regions $S_i$, generally speaking any function $f_{n}(r,\theta)$ that can be decoupled into a direct product of radial and angular functions will have the above property (that ${\langle P_n Y_n \rangle_{S_1}}$ and ${\langle P_n Y_n \rangle_{S_2}}$ always differ by a constant coefficient independent of $Y_n(\theta)$). This means that there may still be room for improvement in finding the optimal post-selection strategy.

\section{Protocol Choice and Key Rate Formula} \label{ref:appendix_rate}

Up to this point, we have obtained simulated statistics $\{\langle Q_k \rangle_{S_{k,i}},\langle QE_k \rangle_{S_{k,i}}\}$ where $k$ is the basis $\{X,Y,Z\}$ and $i$ is the decoy setting (which can be e.g. $i=1,2,3$). We have also described how to use them to bound the single-photon yield and QBER, $Y_1^{k,L},e_1^{k,U}$ for every basis. Note that, up to here, we are still defining a general source capable of sending any polarization state, which we post-select into $H,V,+,-,L$ and $R$ states in the X,Y, and Z bases. We then obtain the statistics with Bob measuring in the X,Y, and Z bases, and the corresponding statistics where Alice sent a single-photon. These data can in principle be used for various protocols, e.g. BB84, RFI-QKD, the three-state (loss-tolerant) protocol, or the six-state protocol.

(1) \textit{BB84 protocol}: We can calculate the key rate of the BB84 protocol using

\begin{equation}
	\begin{aligned}
		R = &P_Z^A P_Z^B \{ \langle P_{1} \rangle_{S_{Z}} Y_1^{Z,L} \left( 1 - h_2(e_1^{X,U})\right) \\
		&-f_e \langle Q_Z \rangle_{S_Z} h_2(\langle QE_Z \rangle_{S_Z}/ \langle Q_Z \rangle_{S_Z}) \}
	\end{aligned}
\end{equation}

\noindent where we assume that $S_{H1}$ and $S_{V1}$ are used for key generation, and $h_2(x)=-xlog_2(x)-(1-x)log_2(1-x)$ is the binary entropy function. Here $P_Z^A$ is the sifting probability for choosing the key generation regions (e.g. $S_{H1}$ and $S_{V1}$), $P_Z^B$ is Bob's basis choice probability (which can take a value close to 1 in the infinite data scenario), $\langle P_1 \rangle_{S_{Z}}$ is the average probability of sending single-photons in the key generation region (in practice we can calculate it by just integrating over $S_{H1}$ since it is symmetric to $S_{V1}$).

(2) \textit{RFI-QKD protocol}: We can alternatively use the data to implement the RFI-QKD \cite{RFIQKD} protocol, which makes use of $XX,XY,YX,YY$ and $ZZ$ statistics for Alice's and Bob's bases. Here we define a rotation-invariable parameter:

\begin{equation}
	C = \sum_{ij=XX,XY,YX,YY} (1-2 e_{1}^{ij,U})^2,
\end{equation}

Following the theory parts of Refs. \cite{RFI_rate1,RFI_rate2}, we define the functions 

\begin{equation}
	\begin{aligned}
		u_{\text{max}} &= min(1,\sqrt{C/2}/(1-e_1^{Z,U})), \\
		v &= \sqrt{C/2 - (1-e_1^{Z,U})^2 u_{\text{max}}^2} / e_{1}^{Z,U},
	\end{aligned}
\end{equation}

\noindent which can be used to bound Eve's information $I_E$ (here $u$ does not need to be optimized but can directly take the value $u_{\text{max}}$ if $e_1^Z \leq 15.9\%$) by:

\begin{equation}
	I_E = (1-e_1^{Z,U}) h_2[(1+u_{\text{max}})/2)] + e_1^{Z,U} h_2[(1+v)/2)].
\end{equation}

We can then obtain the decoy-state RFI-QKD key rate

\begin{equation}
	\begin{aligned}
		R = &P_Z^A P_Z^B [ \langle P_{1} \rangle_{S_{Z}} Y_1^{Z,L} \left( 1 - I_E\right) \\
		&-f_e \langle Q_Z \rangle_{S_Z} h_2(\langle QE_Z \rangle_{S_Z}/ \langle Q_Z \rangle_{S_Z}) ].
	\end{aligned}
\end{equation}

Here we only simulate the above two protocols as examples, but note that the passive source is protocol-independent, and can e.g. be applied to the three-state loss-tolerant protocol, or to the six-state tomographically complete protocol. It is in principle even applicable to e.g. polarization or time-bin phase encoding MDI-QKD too, if Alice and Bob each hold such a passive source.

\section{Finite-Size Effects} \label{ref:appendix_finite_size}

In this Appendix we briefly discuss finite-size effects in the passive scheme. Note that a rigorous finite-size analysis will still be subject to future studies and here we are merely providing an intuition for estimating how much effect a finite data size will have on the key rate. For simplicity, here we only consider collective attacks and assume that all random variables follow a Gaussian distribution. Note that, in principle, we can extend the analysis to coherent attacks by using random sampling and Chernoff's bound \cite{finiteBB84,finiteMDI}.

If we consider only collective attacks and assume that each signal is \textit{i.i.d.}, for a given observable such as the detected counts $Q_{\mu_i}N_{\mu_i}$ (where $Q_{\mu_i}$ is the gain and $N_{\mu_i}$ the number of signals sent in this setting), we can use standard error analysis to approximate the distribution with a Gaussian distribution if the sample size is large, and assume that the standard deviation of $\sqrt{Q_{\mu_i}N_{\mu_i}}$, and the expectation value of the observable $\overline{Q_{\mu_i}N_{\mu_i}}$ can be bounded with

\begin{equation}
	Q_{\mu_i}N_{\mu_i} - \gamma\sqrt{Q_{\mu_i}N_{\mu_i}} \leq \overline{Q_{\mu_i}N_{\mu_i}} \leq Q_{\mu_i}N_{\mu_i} + \gamma\sqrt{Q_{\mu_i}N_{\mu_i}},
\end{equation}

\noindent within a confidence interval determined by $\gamma$, which is the number of standard deviations given a failure probability $\epsilon=2(1-CDF(x+\gamma \sigma))$, where $CDF$ is the cumulative Gaussian distribution and $\sigma$ is the standard deviation of the variable $x$), e.g. for a failure probability of $\epsilon=10^{-7}$, we have that $\gamma\approx 5.3$).

As a simple example to illustrate our idea, let us consider a scenario where Alice randomly chooses between two intensities $\mu_1$ and $\mu_2$ when sending a signal. The asymptotic gain is now $Q{\mu_1\mu_2}=(Q_{\mu_1} + Q_{\mu_2})/2$. Then for each signal, the passing probability is still identical and independent, namely $Q_{\mu_1\mu_2}$. This means that we can simply bound $Q_{\mu_1\mu_2}N_{\mu_1\mu_2}$ with the above standard error analysis too (by approximating the counts with a Gaussian distribution, which is reduced from the binomial distribution of independent and identical ``coin tosses"; it is just that in this case, one first randomly selects among two differently weighted coins, and then performs the toss, but each event is still $i.i.d.$). In fact, this method of bounding the statistical fluctuation of combined data from different intensities has already been used in QKD such as in Ref. \cite{mdiqkd4int} where it is called ``joint-bound analysis".

For the fully passive setup, each fixed setting $(\mu_H,\mu_V,\phi)$ will correspond to a given asymptotic value for $Q$, and the choice of $(\mu_H,\mu_V,\phi)$ follows a classical distribution $p(\mu_H,\mu_V,\phi)$. Overall, the average gain for a given region $\langle Q \rangle_{S_i}$ is the integral (i.e. weighted sum) of a continuous spectrum of settings over the post-selection region $S_i$. However, even in this case, for each individual signal, it simply first picks a ``coin" $Q(\mu_H,\mu_V,\phi)$, and then tosses it, and overall the signals still are \textit{i.i.d.}. Conceptually this is no different from the above two-intensity joint-bound analysis, meaning that we can simply bound $\langle Q \rangle_{S_i} N_{S_i}$ with the same standard error analysis.

\begin{equation}
	\begin{aligned}
		&\langle Q \rangle_{S_i} N_{S_i} - \gamma\sqrt{\langle Q \rangle_{S_i} N_{S_i}} \\
		\leq& \overline{\langle Q \rangle_{S_i}} N_{S_i} \\
		\leq& \langle Q \rangle_{S_i} N_{S_i} + \gamma\sqrt{\langle Q \rangle_{S_i} N_{S_i}}.
	\end{aligned}
\end{equation}

\noindent Note that here the bracket $\langle Q \rangle$ and overline $\overline{Q}$ are used to denote two different averages: the former is the average over possible source settings $(\mu_H,\mu_V,\phi)$ within an acceptable post-selection region $S_i$, while the latter is averaged over independent signals sent over a time interval. The parameter $\langle Q \rangle_{S_i} N_{S_i}$ used here is the simulated experimental observable (while in practice it will be directly obtained from the experiment). $\overline{\langle Q \rangle_{S_i}}$ means that for each individual signal, it has a single observable quantity $\langle Q \rangle_{S_i}$ (meaning it is detected by Bob and it could have come from any setting within the set $S_i$), while this observable quantity averaged over all events to obtain the asymptotic value $\overline{\langle Q \rangle_{S_i}}$. Here the number of signals sent in the given setting is

\begin{equation}
	N_{S_i} = N \iiint_{S_i} p(\mu_H,\mu_V,\phi) d\mu_Hd\mu_Vd\phi,
\end{equation}

\noindent where $N$ is the total signals prepared in the experiment. Note that, for active encoding, the decoy/basis setting should ideally be driven by random number generators, so the number of signals sent in a given setting is a random variable too. However, for the finite-size analysis, we consider $N_{\mu_i} = P_{\mu_i} N$, where $N_{\mu_i}$ is actually the asymptotic (rather than observed) number of signals prepared in a given setting, with which we can obtain the asymptotic gain $\overline{Q_{\mu_i}} = \overline{Q_{\mu_i}N_{\mu_i}}/N_{\mu_i}$. Here similarly for the passive scenario, we should directly use the asymptotic $N_{S_i}$ and should not consider the fluctuations in the actual number of signals prepared in a given setting (i.e. falling within a given post-selection region).

\newpage 
\end{document}